\documentclass{ws-procs961x669}            
\begin{document}
\title{Elementary introduction to Moonshine}

\author{Shamit Kachru}

\address{Department of Physics, Stanford University\\
Palo Alto, CA 94305, USA\\
$^*$E-mail: skachru@stanford.edu}

\begin{abstract}
These notes provide an elementary (and incomplete) sketch of the objects and
ideas involved in monstrous and umbral moonshine.   They were the basis for
a plenary lecture at the 18th International Congress on Mathematical Physics, and
for a lecture series at the Centre International de Recontres Mathematiques school on 
``Mathematics of String Theory."

\end{abstract}


\bodymatter
\bigskip
\noindent
This series of lectures will cover aspects of the subject of ``moonshine,'' and also some related developments in theoretical physics.  
The subject is notable because it ties together -- in an evidently deep way -- widely disparate areas  
of modern mathematics and theoretical physics.  This strongly suggests the existence of an underlying unification of ideas that we are still missing.  A beautiful elementary exposition of some of the same material covered here can be found in the popular book  [\!\!\citenum{Ronen}].

\medskip
The rough plan is to discuss four sets of topics (spread over 5 lectures):

\medskip
\noindent
1.  Monstrous moonshine (which was primarily developed c. 1979-1992; good references
include [\!\!\citenum{FLM,Gannon,Ono}]).

\medskip
\noindent
2.  Mathieu and umbral moonshine (c. 2010 -, with primary references being [\!\!\citenum{EOT,Umbral}]).

\medskip
\noindent
3.  Relations to quantum gravity in 3d (c. 2007 -, with a nice first discussion in [\!\!\citenum{Witten3D}]).

\medskip
\noindent
4.  Enumerative geometry of K3 surfaces and moonshine (a topic under development now [\!\!\citenum{KKP,K3inv}]).

\medskip
\noindent
The first two topics can be considered the basic content of moonshine, while the third and fourth concern
the intersection of ideas developed in the study of moonshine with cognate areas in theoretical physics (quantum gravity and
enumerative geometry of Calabi-Yau manifolds, respectively).
The fourth topic was included in the lecture at the ICMP, but omitted due to constraints of time in the lectures presented in 
Marseilles.  We have included it here for completeness.
In fact, originally, it was also intended to discuss a fifth inter-related subject -- the use of modular
forms associated to 2d CFTs (as partition functions or indices) to provide general insight into which 2d CFTs admit large radius AdS3 gravity duals.  However, this subject is thematically distant enough from the topics covered here that we have not attempted to discuss it.

\medskip
\noindent
The unifying theme of the lectures is the interplay between special symmetries, 2d CFT, modular forms, geometry, and string theory.

\section{Lecture 1: Monstrous moonshine}

\subsection{Numerology}

\subsubsection{The Monster}

Just as we think of the natural numbers as being classified by a prime factorization
$$ n = p_1^{a_1}p_2^{a_2}\cdots $$
it is natural to think of finite groups in terms of ``building blocks."  
It turns out that by defining ``composition series" which use the notion of a normal subgroup, one can
very roughly construct finite groups out of a set of ``primes" -- the finite simple groups (for an 
elementary discussion, see [\!\!\citenum{Ramond}]).  

The classification program for these analogues of primes was completed in the 1980s.  The list includes

\medskip
\noindent
$\bullet$ 18 infinite families (c.f. clock arithmetic with prime clocks)

\medskip
\noindent
$\bullet$ 26 oddballs which do not fit into any of the infinite families.  These are known as ``sporadic
simple groups."  See Figure 1.  The largest of these is the Fischer-Griess Monster $M$.  You can read 
about it in [\!\!\citenum{BorchM}].  It has order:
$$|M| = 2^{46}\cdot 3^{20}\cdot5^9 \cdot 7^6 \cdot 11^2 \cdot 13^3\cdot  17 \cdot 19 \cdot 23 \cdot 29 \cdot 31 \cdot 41 \cdot 47 \cdot 59 \cdot 71 \sim 8 \times 10^{53}$$

\begin{figure}
\label{sporadic}
\begin{center}
\includegraphics[width=0.75\textwidth]{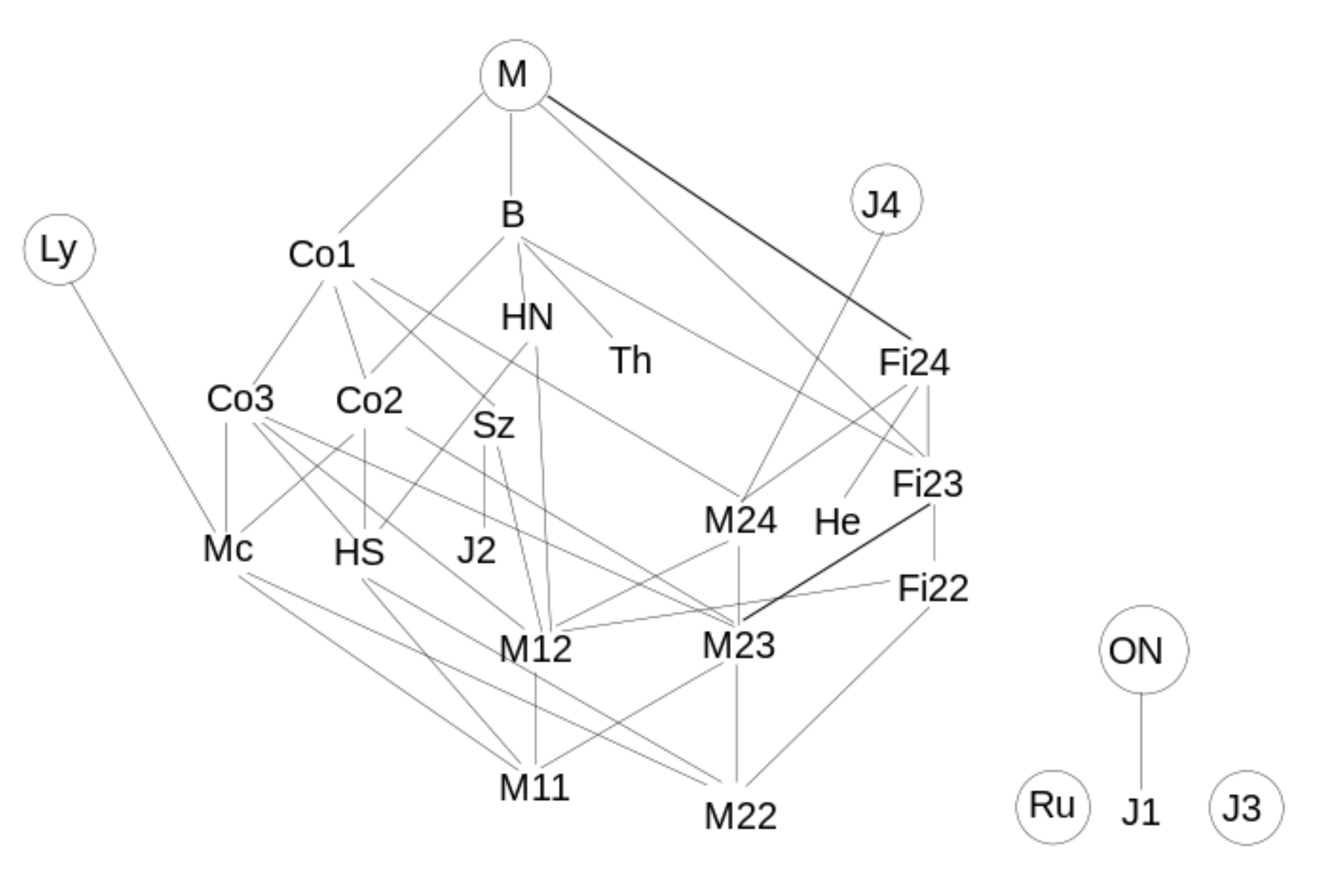}
\end{center}
\caption{The sporadic simple groups.  Figure from [\!\!\citenum{Monsterfig}].
}

\end{figure}

The irreducible representations of the Monster include 
$$\begin{array}{cc}
{\rm Irrep}&{\rm Dimension}\\
\rho_0&1\\
\rho_1&196883\\
\rho_2&21296876\\
\cdots&\cdots\\
\end{array}$$

\subsubsection{The J function}

Modular functions and forms show up in many places in string theory.

Perhaps the most elementary setting is the following. The torus partition function of a 2d CFT with left/right Virasoro generators $L_n, \bar L_n$ and central charge $c$ is
$$Z(\tau,\bar\tau) = {\rm Tr}\left( q^{L_0 - {c\over 24}} \bar q^{\bar L_0 - {c\over 24}}\right),~~q=e^{2\pi i \tau}~.$$
For a chiral theory it is instead
$$Z(\tau) = {\rm Tr}\left( q^{L_0 - {c\over 24}} \right)~.$$
Here, $\tau$ is parametrizing the shape of the torus.
  It is well known that we can parametrize the inequivalent conformal structures by
a complex parameter $\tau$, with $\tau$ naturally taking values in the upper half-plane $H$:
$$\tau = x + iy ~,~~y > 0.$$
We can think of the torus as being generated by the lattice $\langle 1,\tau \rangle$.  

The partition function should behave nicely under large diffeomorphisms of the torus which yield the same
conformal structure.
However there are different choices of generators which yield the same lattice, and hence the
same torus. Choosing
$$\tau^\prime = {{a\tau + b }\over {c\tau + d}}$$
with $$a,b,c,d \in {\bf Z},~~ad - bc = 1$$
yields the same conformal structure on the torus.  There is an $SL(2,{\bf Z})$ freedom here.
The resulting $\tau$ parameter is then naturally taken to live in the {\bf keyhole region} (see Figure 2).
That is, every point in the upper half-plane can be mapped into this region, by a suitable choice of
$SL(2,{\bf Z})$ transformation.

\begin{figure}
\label{keyhole}
\begin{center}
\includegraphics[width=0.95\textwidth]{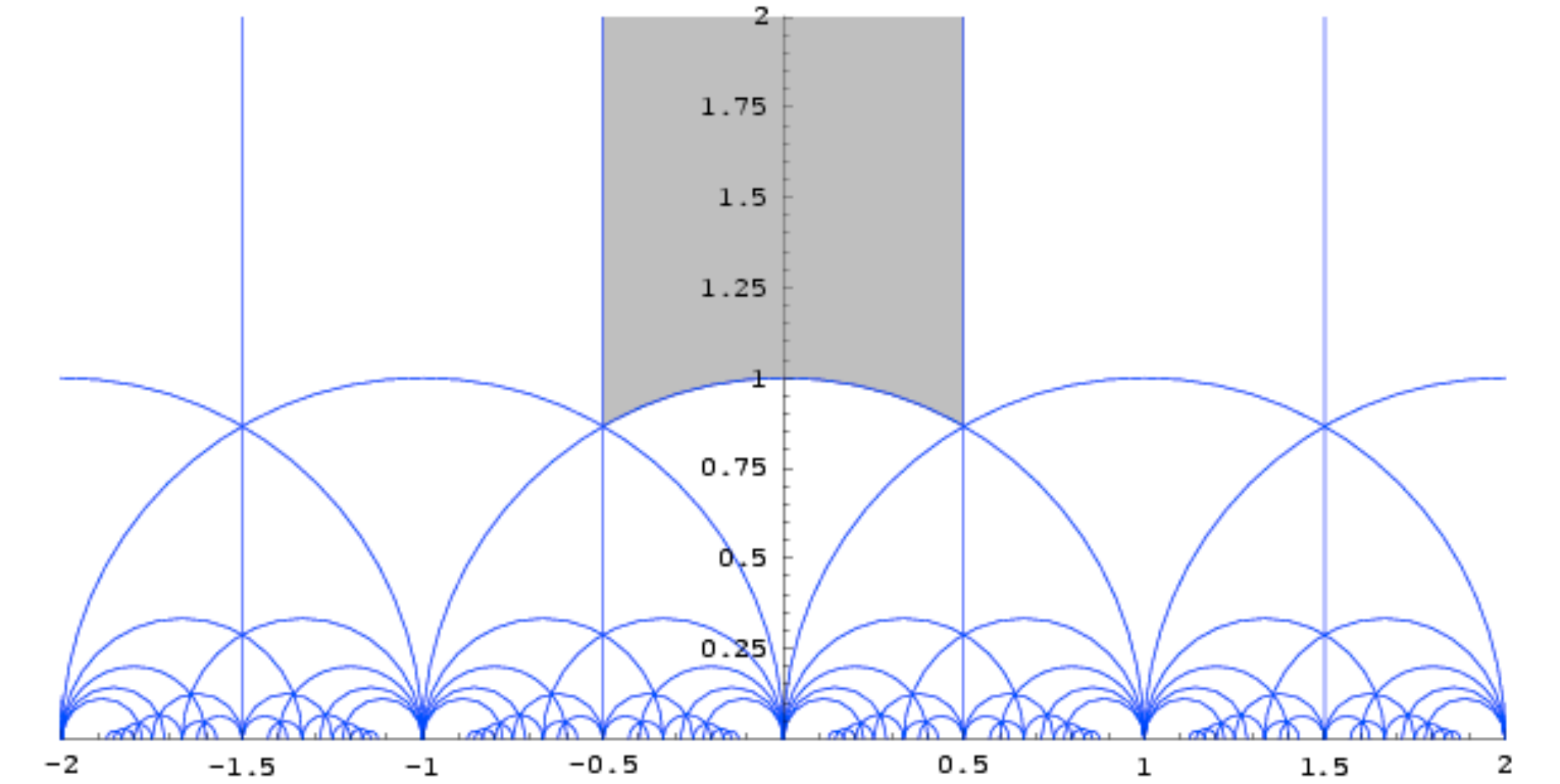}
\end{center}
\caption{Fundamental domain for $SL(2,{\bf Z}) \backslash H$.  Figure from [\!\!\citenum{keyholefig}].
}

\end{figure}

Because of this, the one-loop partition function is naturally viewed not as a general complex function
of $\tau$, but as a modular function:
$$Z({{a\tau + b} \over {c\tau + d}}) = Z(\tau)~.$$
Modular functions are a special case of a more general class of objects, known as modular forms of weight $k$.  Such objects transform
under the modular group as
$$f({{a\tau + b}\over {c\tau + d}}) = (c\tau + d)^k f(\tau)~.$$
A good introduction to the theory of modular forms is [\!\!\citenum{Koblitz}].

Just as one can generate meromorphic functions on the $z-$plane by taking rational functions of $z$, there is a
similar story for modular functions.  The preferred modular function $J(\tau)$ -- the Klein J-function -- maps the keyhole
region bijectively to all of ${\bf C}$.  Then modular functions are rational functions of $J$.

It is natural to expand modular functions in the variable $q$, as $SL(2,{\bf Z})$ is generated by
$$T: \tau \to \tau + 1,~~S:\tau \to -{1\over \tau}~,$$
and this builds in the $T$-invariance.  For the Klein J-function, the resulting expansion takes the form
$$J(\tau) = {1\over q} + 196884 q + 21493760 q^2 + \cdots$$

Looking back at the table of low-lying Monster irreps, it is hard to miss the coincidences:
$$196884 = {\rm dim}(\rho_0) + {\rm dim} (\rho_1)$$
$$21493760 = {\rm dim}(\rho_0) + {\rm dim}(\rho_1) + {\rm dim}(\rho_2)~.$$
What, if anything, do these mean?

\subsubsection{Adding some physics}

Let us try to interpret these findings in a plausible physical setting, with a good dose of hindsight.

We are finding a modular invariant function, with states at a given power of $q$ naturally decomposed into
(at low energies, low-lying) Monster irreps.  Where would we find a modular invariant function with natural
representations of a symmetry group?

In physics, if one has a quantum system with a Hilbert space ${\cal H}$ of physical states, and a partition
function
$$Z(\beta) = {\rm Tr}\left( e^{-\beta H} \right)~$$
(with $\beta$ being the inverse temperature),
then (for integer energy levels) one would have
$$ Z = \sum_n c_n e^{-\beta n} ~.$$
The coefficient $c_n \in {\bf Z}$ counts the number of states at energy level $n$.

Here, we have a modular function -- $J$.  Modular functions arise as torus partition functions in 2d CFT.  
Could we have a 2d (chiral) CFT where
$$J  = {\rm Tr}\left( q^{L_0 - {c\over 24}} \right)~?$$
(To clearly see the analogy with a standard thermal partition function, set $\tau = {i \beta \over 2\pi}$.)
If so, then we should decompose the Hilbert space:
$${\cal H} = \oplus_n {\cal H}_n$$
with 
$$J = \sum j_n q^n,~~{\rm dim} {\cal H}_n = j_n~.$$

If the 2d CFT has Monster symmetry, then it would be natural for the states at a given energy -- spanning  ${\cal H}_n$ -- to transform in interesting Monster representations.  So one would {\bf expect} the $j_n$ to be simple sums of dimensions of
irreducible representations of the Monster.

McKay and Thompson proposed a simple test of this idea.  Suppose the above is true.  I.e., we have some natural
decompositions
$${\cal H}_n = \rho_n = \sum_{i=1}^{k_n} \rho_{n_i}~,$$
with $\rho_n$ the (reducible) representation appearing at the $n$th energy level in the Hilbert
space, and $\rho_{n_i}$ being the irreducible representations appearing in its decomposition.
Then we could compute the McKay-Thompson series:
$$Z_g \equiv {\rm Tr} \left( g q^{L_0 - {c\over 24}} \right)$$
for any $g \in M$.  And we'd know the answer.  If the character of $g$ in the $\rho_{n}$ (generally reducible) representation is
${\rm ch}_{\rho_n}(g)$, then we'd expect
$$ Z_g \equiv {\rm Tr}\left( g q^{L_0 - {c\over 24}} \right) ~=~\sum_{n} ch_{\rho_n} (g) q^{n}~.$$

Some nice facts about $Z_g(\tau)$ and related objects:

\medskip
\noindent
$\bullet$ This function is a class function.  As it is a trace, conjugation $g \to hgh^{-1}$ leaves $Z_g$ invariant.
So the maximal number of distinct functions obtained in this way is equal to the number of conjugacy classes of the Monster -- 194.

\medskip
\noindent
$\bullet$ The trace with a $g$ insertion is not going to be invariant under all of $SL(2,{\bf Z})$.  The torus with
boundary conditions $(f,g)$ around the $a,b$ cycles maps under modular transformation to the torus with boundary conditions $(f^a g^b, f^c g^d)$.\footnote{Here, just for simplicity, I've assumed that $f$ and $g$ commute.}
This is not usually the same as $(f,g)$.  But it will be when the boundary conditions happen to come back to themselves.
For the special case of $(1,g)$ boundary conditions, this yields congruence subgroups $\Gamma_g$ of $SL(2,{\bf Z})$ that will
act nicely on $Z_g$.  In fact, $Z_g$ turns out to be a ${\bf Hauptmodul}$ (this term will be defined carefully below) on $\Gamma_g \backslash  H$.\footnote{This is an oversimplification.
As we describe momentarily, the McKay-Thompson series are typically Hauptmoduln for larger groups which are subgroups of $SL(2,{\bf R})$ but
not $SL(2,{\bf Z})$.  This extension of the symmetry is not manifest from general considerations of conformal field theory.}

\medskip
This in turn implies some interesting and non-trivial checks we can make given any guess about
how the ${\cal H}_n$ decompose into irreducible representations of $M$.  Because for a given a priori decomposition of the chiral CFT statespace
into monster irreps, we can compute the twinings either via the formula above, or by using the fact that it is modular
under $\Gamma_g$ (which gives the full answer after a finite amount of linear algebra).  

\medskip
\noindent
$\bullet$ Related fact: there is a deep tie between Monstrous moonshine and genus zero function fields.  
If we consider the keyhole region $SL(2,{\bf Z}) \backslash H$ for instance, after adjoining the point $\tau = i \infty$,
we can think of this space as a compact Riemann surface.  The modular function $J$ maps $SL(2,{\bf Z}) \backslash H$ in a
one to one, onto way to ${\bf C}\cup \infty$.  This shows that the Riemann surface associated to $SL(2,{\bf Z})\backslash H$
is of genus zero.  So, the field of meromorphic functions on this Riemann surface is given by all rational functions
of $J$ with complex coefficients:
$$k = {\sum_{i=0}^{n} a_i J^i \over \sum_{i=0}^{m} b_i J^i}~.$$
This is called a genus zero function field, and the generator $J$ is called a Hauptmodul.

\medskip
We can repeat this logic for other subgroups $\Gamma \in SL(2,{\bf R})$ of genus zero; i.e the groups which
share the property that $\Gamma \backslash H$ yields a Riemann surface of genus zero.  The field of rational functions
on $\Gamma \backslash  H$ is again a genus zero function field, with a generator $j_{\Gamma}$ analogous to $J$.  Examples
of genus zero subgroups of $SL(2,{\bf R})$ are the congruence subgroups $\Gamma_0(p)$ with $p$ prime and
$p-1 | 24$.  Here
$$\Gamma_0(N) \equiv \{ \left(\begin{array}{cc}a&b\\c&d\end{array}\right) \in SL(2,{\bf Z}) | c \equiv 0 ~{\rm mod}~N
\}~.$$

\medskip
A group of greater interest here is $\Gamma_0(N)^+$.  This group is obtained by adding the Fricke involution
$$ \tau \to - {1\over N\tau} $$
as an additional generator to $\Gamma_0(N)$.  It is a beautiful and mysterious fact that the primes $p$ for which $\Gamma_0(p)^+$ has
genus zero, are precisely the primes that divide the order of the Monster!\footnote{For a physical explanation of the enhancement of the modular group to include Fricke involutions, see [\!\!\citenum{NatRob}].}

\medskip
\noindent
$\bullet$ There is a slightly generalized story, called (creatively) ``generalized moonshine," studying the
partition functions $Z_{f,g}$ with twists by $f$ and $g$ on the spatial and temporal circles of the Euclidean
torus [\!\!\citenum{generalized}].  These functions have beautiful modular properties as well, as expected.
Rigorous discussion of the ideas behind generalized moonshine can be found in [\!\!\citenum{Carnahan}].

\subsection{Constructing the moonshine module}

In fact, at a physicists level of understanding, the direct construction of a 2d CFT elucidating this connection was achieved in the mid 1980s.  Frenkel, Lepowsky, and
Meurman realized that one can construct a CFT giving rise to ${\cal H}$ as follows [\!\!\citenum{FLM}].

Consider an even, self-dual,  Euclidean lattice $\Lambda$ of dimension $r$.  It gives rise to a consistent
$c=r$ holomorphic CFT.  To discuss this, introduce $r$ fields $X^i(z)$, viewed as coordinates on
$$T^r = {{\bf  R}^r \over \Lambda}~.
$$
We could of course consider lattices that aren't even, self-dual lattices, but the statement
(not hard to prove, but we won't here) is that they won't (by themselves) lead to modular invariant
partition functions.

\medskip
Now even, self-dual lattices of low dimension are rare objects.  They occur only in 8k dimensions.
For a discussion of their properties, see e.g. [\!\!\citenum{Conwaybook}].

\medskip
\noindent
The {\bf number} of such lattices in a given dimension is interesting:
$$\begin{array}{cc}
{\rm dim}&{\rm Number}\\
8&1\\
16&2\\
24&24\\
32&{\rm a~whole~lot}
\end{array}$$

\medskip
In all dimensions $\geq 32$, the number is huge and grows very quickly.  Bounds on the number can
be obtained by using the Siegel mass formula (see e.g. [\!\!\citenum{Conwaybook}]).  The low-dimensional examples are all interesting and useful in physics.  In dimension 8, we find the $E_8$ root lattice.  Dimension 16 sees both the 
$SO(32)$ lattice and $E_8 \times E_8$ -- both of which play crucial roles in heterotic string theory.
The 24 examples in dimension 24 will appear prominently in the sequel.

\medskip
\noindent
Among  the 24 in dimension 24, particularly interesting is the Leech lattice [\!\!\citenum{Leech}].  It gives the densest way a grocer could pack oranges in 24-dimensional space [\!\!\citenum{oranges}].\footnote{A nice pedagogical discussion of lattices and sphere packings,
and their connections to codes and simple finite groups, can be found in [\!\!\citenum{Thompson}].  I thank Pierre Ramond for introducing me to this excellent little book.}

It is a fact of life that
$$
Z_{\rm Leech}(q) = {\Theta_{\rm Leech}\over \eta^{24}} = {1\over q} + 24 + 196884 q + \cdots
$$
$$= J(q) + 24.
$$
The $\Theta$ function gives the sum over the momenta / windings in the lattice, while the $\eta$
functions come from the bosonic oscillator modes.

As {\bf all} of the 24 even unimodular positive lattices give $J(q) + {\rm const.}$ for their partition function,
we can understand the computation of the 196884 in many different ways.  Easier than actually going through the
computation for the Leech lattice, is doing it for $E_8^3$.  So, lets do that.

The $E_8$ root lattice is 
$$\Gamma_8 = \{ x_i \in {\bf Z}^8 \cup ({\bf Z}+1/2)^8:
\sum_i x_i \equiv 0 ~({\rm mod})~2 \}~.$$
The lattice theta function
$$\Theta_{E8} = \sum_{{\bf v}\in \Gamma_8} q^{{1\over 2} {\bf v} \cdot {\bf v}} = {1\over 2}\left( \theta_2(q)^8 + \theta_3(q)^8 + \theta_4(q)^8\right)$$
$$ = 1 + 240 q + 2160 q^2 + \cdots$$
Here $\theta_i$ are the Jacobi theta functions, whose q-expansion can be found in
any standard reference on modular functions.

So taking three copies, the partition function of the CFT
of chiral bosons propagating on the $E_8$ lattice will 
have terms at $O(q)$ that arise as:

$$O(q):~3 \times 240^2 + 3 \times 2160 + 3 \times 24 \times 240 + 324 ~=~196884~.$$

\noindent
In understanding the origin of these terms, it is important to remember the factor of  
$${1\over \eta(q)^{24}} = {1\over q} \left(1 + 24 q + 324 q^2 + \cdots\right)~.$$

\medskip
Let us return again to the chiral CFT based on the Leech lattice.  Its Hilbert space is
{\bf almost} the one we want to furnish a possible realization of the moonshine module.
Frenkel, Lepowsky, and Meurman proved that:

\medskip
\noindent
$\bullet$ An appropriate ${\bf Z}_2$ quotient of this theory removes the pesky `24' in the
partition function, yielding a quotient theory with partition function $J(q)$ on the nose.

\medskip
\noindent
$\bullet$ It also gives a theory where $M$ acts as a symmetry, commuting with the Hamiltonian
and leaving the OPEs unchanged.  This provides a principled derivation of decompositions of each coefficient
in the $q$-expansion of the $J$-function, into sums of dimensions of irreducible representations of the Monster.

\medskip
\noindent
The existence of this CFT provides a heuristic explanation for some aspects of Monstrous moonshine, while leaving others -- like the genus zero property
of the McKay-Thompson series under suitable subgroups of $SL(2,{\bf R})$ - shrouded in mystery.
The mathematically rigorous understanding due to Borcherds required the introduction of highly nontrivial new ideas and identities.  It is nicely discussed in his informal write-up of his ICM lecture [\!\!\citenum{Borcherds}].

\medskip
Where do we go next?
We plan to deepen and extend our understanding in a few directions.

\medskip
\noindent
$\bullet$ A new class of moonshines was discovered through the study of the simplest Calabi-Yau compactificaton (on $K3$) in 2010.  This will be our subject next time.  The 24 interesting lattices in dimension 24 will reappear in Lecture 3.

\medskip
\noindent
$\bullet$ The Monster CFT reappears in a natural role when one considers AdS3/CFT2 duality.
We will discuss this connection in Lecture 4.

\section{Lecture 2: Mathieu moonshine}

\bigskip
\noindent
In the first lecture, we witnessed a remarkable relationship discovered through a coincidence between the q-expansion of the J function and the dimensions of the low-lying irreducible representations of the Monster group.  This presaged a relationship between group theory and modular forms (mediated via string theory) that remains mysterious till the present day.

Today, we add two more ingredients - algebraic geometry and mock modular forms.

\subsection{K3 and its supersymmetric index}

The compactification of string theory on Calabi-Yau manifolds is of tremendous interest as a ``philosophical tool."
Recall that a Calabi-Yau space is a complex K\"ahler manifold of vanishing first Chern class.  Calabi conjectured, and Yau proved, that such a space admits a Ricci flat metric for each choice of the K\"ahler class.  The tie to string theory comes about through the connection between Ricci flatness and vacuum Einstein equations -- Ricci flat manifolds solve the vacuum Einstein equations.
The Calabi-Yau spaces have the additional merit that they admit convariantly constant spinors, and
so preserve some fraction of the space-time supersymmetry.
Uses that strings on these spaces have enjoyed include:

\medskip
\noindent
$\bullet$ model building (see e.g. [\!\!\citenum{Uranga}])

\medskip
\noindent
$\bullet$ geometric engineering of soluble field theories (see e.g. [\!\!\citenum{geomeng}])

\medskip
\noindent
$\bullet$ computations of black hole entropy (see e.g. [\!\!\citenum{stromingervafa}])

\medskip
\noindent
$\bullet$ deep ties to enumerative geometry (see e.g. [\!\!\citenum{candelas}])

\medskip
\noindent
A highly readable popular-level discussion of Calabi-Yau compactification and its many uses in theoretical physics appears in [\!\!\citenum{Yaubook}].

\bigskip
The simplest non-trivial example of a Calabi-Yau space is the $K3$ surface.
A concrete hypersurface which is topologically a $K3$ and which admits a Ricci flat metric is
the quartic in ${\bf P^3}$
$$ \sum_{i=1}^{4} z_i^4 = 0~,$$
for instance.
  Type IIA string theory on K3 has an 
80-dimensional moduli space of vacua [\!\!\citenum{AM}]
$${\cal M}_{K3} = O(4,20;{\bf Z}) \backslash O(4,20) \slash (O(4) \times O(20))~.$$
For an excellent review with a comprehensive discussion of the physics and mathematics of strings on K3,
see [\!\!\citenum{Aspinwallreview}].

At most points in this moduli space, we certainly cannot ``solve" the CFT.  In particular, we can't compute the partition
function $Z$.  This is not surprising.  Yau's theorem guarantees existence of a Ricci flat metric, for instance,
but does not construct it for us.  At large radius, where this metric would be a good approximation to the relevant
CFT data, we therefore can't compute $Z$ without first solving this (highly non-trivial) problem -- as even for a 
point particle theory, computation of the partition function would require knowledge of the metric on the space
where the particle lives.  No exact Ricci-flat metric on a smooth, compact Calabi-Yau space has been determined to date.  For recent progress
in numerically approximating such metrics, see e.g. [\!\!\citenum{Headrick, Mike}].

So, the upshot is, we cannot compute the partition function.
However, life is kind to us in allowing us to discuss supersymmetric indices instead.

\subsubsection{Index kindergarten}

The simplest supersymmetric index is the Witten index [\!\!\citenum{Edindex}].  Suppose one has a family of supersymmetric quantum mechanics
theories, living on a moduli space ${\cal M}_{SQM}$.  The Hamiltonian is
$$H = \{ Q, Q^\dagger \}$$
where $Q$ is the supercharge, with
$$Q^2 = 0~.$$
There is also a ${\bf Z}_2$, $(-1)^F$, under which $Q$ is odd.

Because this simplest example of a
Clifford algebra has very simple representation theory -- one dimensional representations
at $H=0$ and two dimensional representations at $H > 0$ -- some important
conclusions immediately follow:

\medskip
\noindent
$\bullet$ States $\vert \alpha \rangle$ with 
$$Q |\alpha \rangle ~=~0$$
are supersymmetry preserving ground states.  They have precisely zero energy.  They may
have either eigenvalue of $(-1)^F$.  Even/odd states are called `bosons' and `fermions,'
respectively.

\medskip
\noindent
$\bullet$ States with $E>0$ are {\bf paired} by the action of $Q$.  There is one boson and one fermion in 
each pair.

\bigskip
Now, imagine moving around in the space of theories, ${\cal M}_{SQM}$.  The finite energy pairs generally move around in energy.  
A boson may leave zero energy, but only if it is accompanied by a fermion -- as otherwise it would furnish an 
unpaired finite energy state.  A boson may come down and join zero energy, but only if it is accompanied by its
fermi partner - for otherwise, that fermion would now furnish an unpaired finite energy state.

So
$${\rm Tr}\left( (-1)^F e^{-\beta H} \right)$$
is a {\bf constant} on ${\cal M}_{SQM}$, governed by the difference of the number of zero energy bosons and
fermions.  (It is important in making this argument precise that we assume that the theory has a discrete spectrum;
otherwise, there can be subtleties which invalidate the argument [\!\!\citenum{Akhoury}].)

The use of this is the following.  Although ${\cal M}_{SQM}$ may be a large space, with Hamiltonians that are completely intractable at most points, perhaps at a handful of points $H_{\cal M}$ simplifies.  We can compute the index there, and be confident that it will remain the same at the generic, intractable points in theory space.

\subsubsection{Indices in 2D}

Now, we promote our discussion to 2D field theory.  Let us imagine our object of interest is a (2,2) supersymmetric
CFT with Calabi-Yau target.  (The K3 CFT is a special example with {\bf even more} supersymmetry).

The (2,2) algebra comes with a $U(1)$ current of each chirality, in addition to the stress tensors and supercharges.
It is then natural to define partition functions and indices flavored by the $U(1)$ quantum numbers.

This allows us to define a much more refined analogue of the index of supersymmetric quantum mechanics, which associated a number to each moduli space ${\cal M}_{SQM}$.  Here instead, we can associate a full modular form to each Calabi-Yau
moduli space.  This modular form is called the {\bf elliptic genus}.  It is defined as [\!\!\citenum{EG}]
$$\phi(\tau,z) = {\rm Tr}_{R,R}\left( q^{L_0-{c\over 24}} y^{J_L} (-1)^{F_R} {\bar q}^{\bar L_0 - {c\over 24}}\right)~.$$
Here
$$q = e^{2\pi i \tau},~~y = e^{2\pi i z}$$
are parameters tracking the (left-moving) energy and $U(1)$ charge of the contributing states.  
The trace is computed in the Ramond sector for both left and right movers.
The
$Tr((-1)^{F_R} \bar q^{\bar L_0})$ is morally a right-moving Witten index, receiving contributions only from
zero energy states.  So the theory (as long as it has discrete spectrum) localizes on right-moving Ramond ground states.

This object has nice modular properties.  It is what is known as a ``weak Jacobi form." 
Weak Jacobi forms are characterized by a weight $w$ and an index $m$.  
They have the properties
$$\phi\left({{a\tau +b} \over {c\tau + d}},{z \over {c\tau + d}}\right) ~=~(c\tau + d)^w e^{2\pi i m {cz^2 \over {c\tau + d}}} \phi(\tau,z)$$
$$\phi(\tau,z + \ell \tau + \ell^\prime) = 
e^{-2\pi i m(\ell^2\tau + 2\ell z)} \phi(\tau,z)~.$$ 
It is a fact of life that the elliptic genus of a Calabi-Yau (complex) $n$-fold is a weak Jacobi form of weight 0
and index $n/2$ [\!\!\citenum{Gritsenko}].  Basic facts about such forms can be found in [\!\!\citenum{Eichler}], and a comprehensive review of their properties and recent uses in theoretical physics appears in [\!\!\citenum{DMZ}].

The space of holomorphic modular forms of weight $w$ has a simple structure:
at a given $w$, it is a vector space with basis given by monomials in the Eisenstein series $E_4, E_6$ of
appropriate weight.  (For example, at weight 12, there are two basis forms, which can be taken as $E_4^3$ and $E_6^2$.)
  The Jacobi forms enjoy a similar story.  Here, the generators are
again the Eisenstein series $E_4, E_6$ as well as the two new generators
$$\phi_{-2,1}(\tau,z) = {\theta_1(\tau,z)^2 \over \eta^6}$$
$$\phi_{0,1}(\tau,z) = 4 \left( {\theta_2(\tau,z)^2 \over \theta_2(\tau,0)^2} + {\theta_3(\tau,z)^2 \over \theta_3(\tau,0)^2} + {\theta_4(\tau,z)^2 \over \theta_4(\tau,0)^2}\right)~,$$
where again the $\theta_i$ are the standard Jacobi functions.
The subscripts denote the weight and index. ($E_4, E_6$ clearly have index 0).

For Calabi-Yau 2-folds, the story is very simple.  The K3 surface must yield a Jacobi form of weight 0 and index 1.
There is one possibility, up to scale -- $\phi_{0,1}$.  In fact, it turns out that with these standard definitions,
$$\phi_{K3}(\tau,z) = 2\phi_{0,1}~.$$
This was computed directly in various CFTs realizing K3 compactifications of string theory in [\!\!\citenum{EOTY}].

\subsubsection{Character decomposition}

Here, we will be even a little bit more sloppy.  But we will try to uncover the basic ideas and spirit of the subject clearly.
Since it is a Jacobi form, one can of course expand the elliptic genus in powers of $q,y$.  I.e. find all the coefficients in 
$$\phi(\tau,y) = \sum_{n,l} c(n,l) q^n y^l~.$$
This will be an interesting expansion for us in lecture 4.  But for today, we'd like to do something a bit
more thoughtful.

The K3 CFT has $N = (4,4)$ supersymmetry.  
In addition to the Virasoro generators and 4 supercharges, each copy of the 2D $N=4$ superconformal algebra comes with an $SU(2)$ R-current.
The representations of $N=4$ superconformal symmetry in 2D
are labelled by a weight $h$ and a (Cartan of $SU(2)$) $U(1)$ charge $J_3$. 
The resulting highest weight representations then have known superconformal characters.
You can look up the precise formulas
in the literature [\!\!\citenum{Taormina}].  The basic point is that there are two types of
characters -- a small number of massless (or BPS) characters, and an infinite tower of massive characters -- and one can expand
$$\phi_{K3}(\tau,y) = ({\rm massless ~characters}) + \sum_{\rm massive} A_n ~ch_{n}(\tau,y)$$
where $ch_n$ is the appropriate charater for a $\Delta \sim n$ super-Verma module with the fixed $U(1)$ charge
allowed by the $c=6,$ $N=4$
symmetry.

The resulting ${\bf numbers}$ are
$$A_1 = 90 ~(= 45 + {\overline {45}}),~~A_2 = 462 ~(= 231 + {\overline {231}}),~~A_3 = 1540 ~(= 770 + {\overline {770}}), \cdots$$

It was noticed by Eguchi-Ooguri-Tachikawa in 2010, that these numbers constitute dimensions of representations
of the sporadic simple group $M_{24}$ [\!\!\citenum{EOT}].

\subsection{Introducing the largest of Mathieu's groups...}

$M_{24}$ is a sporadic group of order
$$|M_{24}| = 2^{10} \cdot 3^3 \cdot 5 \cdot 7 \cdot 11 \cdot 23 ~=~244,823,040~.$$
It is a subgroup of the permutation group on 24 objects.

\medskip
If you look this thing up in Wikipedia, what you learn is that it is the ``automorphism group of the unique doubly even self-dual code of length 24 with no words of length 4 (also called the extended binary Golay code)."

\medskip
Here is a perhaps more friendly un-packaging of that description.

\medskip
\noindent
$\bullet$ Consider a codeword comprised of 24 0s and 1s.

\medskip
\noindent
$\bullet$ Any length 24 codeword has even overlap
with all codewords in $G$ iff it is in $G$.

\medskip
\noindent
$\bullet$ The number of 1s in each codeword is divisible
by 4, but not equal to 4.

\medskip
\noindent
$\bullet$ Then, the subgroup of $S_{24}$ that preserves
$G$, is $M_{24}$.

\medskip
The group $M_{23}$ (also sporadic simple), can be defined as a subgroup of $M_{24}$ that preserves a point.
For more details about these groups, see e.g. [\!\!\citenum{Conwaybook}].

\subsubsection{Why is $M_{24}$ appearing here?}

This is not well understood, though I will present some recent work which I think sheds light on this, in Lecture 3.

Here are some relevant observations:

\medskip
\noindent
$\bullet$ It is known from work of Mukai and Kondo that the symplectic automorphisms of any $K3$ surface
(symmetries which preserve the holomorphic (2,0) form) lie in subgroups of $M_{23}$ [\!\!\citenum{Mukai}].
But the groups that are attained are not so huge -- being proper subgroups of $M_{23}$ of relatively low order.

\medskip
\noindent
$\bullet$ Then, one's natural thought is: can `stringy' effects enhance the relevant group to $M_{24}$?
Unfortunately, it is known from work of Gaberdiel-Hohenegger-Volpato [\!\!\citenum{GHV}] that no CFT with $K3$ target admits
$M_{24}$ symmetry, and not all symmetries of K3 CFTs fit in $M_{24}$.  They do, however, fit in $Co_0$ -- a double cover
of Conway's largest sporadic group, $Co_1$.\footnote{It so happens that $Co_0$ is the symmetry group of the Leech lattice, a fact which will play
an important role in later lectures.  An enjoyable biographical account of Conway's career can be found in [\!\!\citenum{Conwaybio}].}
In fact, subgroups of $Co_0$ which preserve a 4-plane in the defining representation, the ${\bf 24}$, are the ones which appear.
(The relevance of choosing 4-planes in a 24-dimensional space can be seen, from various viewpoints relevant to K3 conformal field theory,
in [\!\!\citenum{AM,GHV,four-planes,K3inv}].)

\medskip
\noindent
We conclude that if there is an explanation of Mathieu moonshine in the elliptic genus of K3, it will not admit a simple analogy with Monstrous moonshine where a specific K3 CFT furnishes the relevant statespace.

\subsection{Mock modular forms}

This new kind of moonshine seems to enjoy two differences from Monstrous moonshine and its closest relatives.

\medskip
\noindent
1) There is a role for algebraic geometry -- visible in the presumed significance (?) of the K3 surface.

\medskip
\noindent
2) Instead of just modular forms, the ``mock modular forms" of Ramanujan-Zwegers also appear.\footnote{
A simple introduction to these appears in [\!\!\citenum{Folsom}], while you can find some more details in 
[\!\!\citenum{Zwegers,Mock}].   
As a cultural aside, for a beautiful biography of Ramanujan, including description of his prescient work on modular forms, see [\!\!\citenum{Kanigel}].}

\medskip
Let us expand on this second point.

\medskip
\noindent
In the character expansion of the elliptic genus, we naturally encounter the function
$$H(\tau) = 2q^{-1/8}\left(-1 + 45 q + 231 q^2
+ 770 q^3 + \cdots \right)$$
This is the function whose $q$ expansion is encoded in the coefficients $A_n$ we described before.
(It is natural that $A_n$ multiplies $q^n$ up to a shift, because the difference between massive characters
of dimension $\Delta$ and $\Delta + 1$ is just an overall power of $q$ -- so one can re-write the sum
over massive characters by pulling out one massive character and then writing a q-expansion with coefficients
governed by the $A_n$).
This function $H$ is a particular example of a more general
class of objects, known as {\bf mock modular forms}.

\medskip
\noindent
The slogan with mock modular forms is that they arise when an object has two choices: to be modular but non-holomorphic, or to be holomorphic but not modular.
The most familiar example of such an object is the
Eisenstein series $E_2$, which is a quasi-modular form:
$$E_2(\tau) = 1 - 24 \sum_k \sigma_1(k) q^{2k}$$
(where $\sigma_1(k)$ is the sum of the positive integer divisors of k).
$E_2$
is holomorphic, but not modular.
$$\hat E_2 = E_2(\tau) - {3\over \pi y}$$
on the other hand, is modular, but not holomorphic.

The more general story involves a holomorphic function
$f(\tau)$ together with a modular ${\bf shadow}$ $s(\tau)$, such that
$$\hat f(\tau,\bar\tau) = f(\tau) + \int_{-\bar \tau}^{\infty}(\tau^\prime + \tau)^{-k} \bar s(-\bar \tau^\prime) d\tau^\prime$$
is modular.  If $\hat f$ has weight $k$, then the shadow $s$ should have weight $2-k$.
 In our simple example -- $E_2$ -- the shadow is a constant.

One can ask ``why" mock modular forms arise in Mathieu moonshine.  At a technical level, the reason is that this moonshine
appears in the character expansion of a conformal field theory with extended supersymmetry, and superconformal characters are characterized
by mock modular behavior [\!\!\citenum{EguchiHikamiMock}].  It is also true that sigma-models with non-compact targets naturally exhibit mock modular behavior in 
their partition functions
[\!\!\citenum{NCchar}]. 
Interestingly, one can also find certain indices (helicity supertraces) in conformal theories related to K3, which give rise to 
the same mock modular forms [\!\!\citenum{HarveyMurthy}].
 We 
expect that a deeper understanding of the appearance of mock objects will emerge with further research.

\section{Lecture 3: Umbral moonshine}
\bigskip
Let me begin with a review of what we saw in the last lecture.

\medskip
\noindent
* We discussed sigma models with K3 target.  They come in a large moduli space
$${\cal M}_{K3} = O(4,20;{\bf Z} \backslash O(4,20) \slash (O(4) \times O(20)~.$$
At generic points in this moduli space, we do not have good control over the theory.
For instance, we cannot compute the partition function $Z(\tau,\bar\tau)$.

\medskip
\noindent
* However, we can compute a 2d supersymmetric index, the elliptic genus.  We 
saw that for $K3$ compactification, simple logic tells us that the answer is that
$$\phi_{K3}~=~{\rm Tr}_{R,R}\left( (-1)^{F_L + F_R} q^{L_0-{c\over 24}} y^{J_L} \bar q^{\bar L_0- {c\over 24}} \right)~ = ~2\phi_{0,1}(\tau,z)~.$$

\medskip
\noindent
* The character expansion of the elliptic genus naturally produces coefficients which have simple decompositions
into dimensions of irreducible representations of the sporadic simple group $M_{24}$.  They are also, in fact, coefficients in the q-series
of a mock modular form.  The role of $M_{24}$ remained mysterious; it is not a symmetry of any K3 conformal
field theory.

\medskip
With that bit of review, we are ready to venture onward.

\medskip
Starting with Cheng, soon after the work of Eguchi-Ooguri-Tachikawa, various groups began to try to precisely define the predictions of
Mathieu moonshine by studying `twined' elliptic genera (genera with an insertion of a symmetry transformation in the trace) [\!\!\citenum{Twinings}].  These serve as the
generalization of the McKay-Thompson series of Monstrous moonshine to this new instance of moonshine.  Roughly, the idea is that one should:

\medskip
\noindent
$\bullet$ Pick an explicit K3 CFT and a symmetry $g$ of order $N$, where $N$ matches the order of some $M_{24}$ conjugacy class $[\alpha]$.  

\medskip
\noindent
$\bullet$ Compute $$\phi_{K3,g} = {\rm Tr} \left( g (-1)^{F_L + F_R} q^{L_0-{c\over 24}} y^{J_L} \bar q^{\bar L_0-{c\over 24}} \right)$$
explicitly.  This is possible for toroidal orbifolds, Gepner models, or more general Landau-Ginzburg orbifolds using slightly fancier
technology [\!\!\citenum{GHV,Roberto,four}].

\medskip
\noindent
$\bullet$ Compare to the would-be twining you would get by identifying $[g]$ with $[\alpha]$, and assuming a given decomposition of the coefficients in
the elliptic genus into irreducible representations of $M_{24}$.  Recall if
$$A_n = {\rm dim} V_n, V_n = \sum_i \rho^n_{i}$$
with $\rho^{n}_i$ being the $M_{24}$ irreducible representations that occur in the guess
for the nth coefficient, then you'd get something like
$$\phi_{K3,g} ~?=?~ \sum_n ch_{V_n}(\alpha) ch_{n}(\tau,z)~.$$

\medskip
\noindent
$\bullet$ The upshot of this research was that these twining functions ``work" for many $g$ with the
right orders, with a suitable choice of decompositions of the statespaces $V_n$ into $M_{24}$ representations.  But there exist some $g$ for which the
twinings do not match those of any $M_{24}$ element
with the same order, and other $g$ which cannot even
hypothetically fit into $M_{24}$.\footnote{A precise prescription of the full predictions of Mathieu moonshine for twinings, with a proposed analogue of the
`genus zero property' of Monstrous moonshine, was put forward in [\!\!\citenum{ChengDuncan}].}

\medskip
$\noindent$ All symmetries $g$ of the K3 CFT do fit into $Co_0$.  Is there a unifying richer story?

\subsection{Umbral moonshine}

In fact, Mathieu moonshine is one of a family of 23 moonshines.  The larger structure was uncovered by
Cheng-Duncan-Harvey [\!\!\citenum{Umbral}], and has been called ``umbral moonshine." (``Umbral" from the latin ``umbra," meaning ``shadow").

\medskip
Recall from Lecture 1 that even, self-dual positive lattices in $d=24$ are rare objects.  There are precisely 24
of them.  One of them is the Leech lattice, of orange packing and moonshine fame.  The others, numbering 23, 
are known as the  ``Niemeier lattices."

\medskip
Each of the Niemeier lattices is canonically associated to an A-D-E root system.  The rules are:

\medskip
\noindent
* One can use any of $A_n, D_n, E_{6,7,8}$.  

\medskip
\noindent
* Each factor appearing in a given lattice must have the same Coxeter number.  
(For a Lie algebra, the dimension is given by $n(h+1)$ where $n$ is the rank and $h$ is the Coxeter number).
A table of ranks and Coxeter numbers appears below.

\medskip
$$\begin{array}{ccc}~&n&h\\
A_k&k&k+1\\
D_k&k&2k-2\\
E_6&6&12\\
E_7&7&18\\
E_8&8&30\end{array}$$

\medskip
\noindent
* The ranks must add up to $24$.

\medskip
Proceeding with these rules, we are able to enumerate the 23 Niemeier lattices.

$$\begin{array}{cc}
{\bf Niemeier ~root ~system}&{\bf Umbral ~symmetry~ G}\\
A_1^{24}&M_{24}\\
A_2^{12}&2.M_{12}\\
A_3^8&2.AGL_3(2)\\
A_4^6&GL_2(5)/2\\
A_5^4D_4&GL_2(3)\\
A_6^4&SL_2(3)\\
A_7^2D_5&{\rm Dih}_4\\
A_8^3&{\rm Dih}_6\\
A_9^2D_6&{\mathbb Z}_4\\
A_{11}D_7E_6&{\mathbb Z}_2\\
A_{12}^2&{\mathbb Z}_4\\
A_{15}D_9&{\mathbb Z}_2\\
A_{17}E_7&{\mathbb Z}_2\\
A_{24}&{\mathbb Z}_2\\
D_4^6&3.{\rm Sym}_6\\
D_6^4&{\rm Sym}_4\\
D_8^3&{\rm Sym}_3\\
D_{10}E_7^2&{\mathbb Z}_2\\
D_{12}^2&{\mathbb Z}_2\\
D_{16}E_8& \\
D_{24}& \\
E_6^4&GL_2(3)\\
E_8^3&{\rm Sym}_3
\end{array}$$

\medskip
\noindent
Shown in the table is also a symmetry associated to each lattice $L$ which we can call the ``umbral symmetry"
$G_L$.  It is the automorphism group of the associated lattice, modulo the Weyl group of the associated A-D-E system.  More precisely, if we let $W^L$ be the normal subgroup of
${\rm Aut}(L)$ generated by reflections in the roots, then the
umbral group
$$G_L := {\rm Aut(L) \over W^L}~.$$

\medskip
It is important to keep in mind that while the Niemeier lattices are based on A-D-E root systems, they are in most cases 
modifications of these including extra ``gluing vectors."  This both allows them to satisfy the even self-dual condition,
and breaks some of the naive symmetry.
For instance, importantly, the $A_1^{24}$ theory does not have $S_{24}$ permutation symmetry, but instead
enjoys only an $M_{24}$ symmetry.

\medskip
The umbral moonshine conjectures of Cheng-Duncan-Harvey amount to the statement that there is a moonshine relating
the Niemeier lattices and their associated symmetry groups $G$, to a vector-valued mock modular form.
If $m$ is the Coxeter number of each A-D-E factor in $L$, then the mock modular form is a vector of $2m$ components (which, in general,
mix under modular transformations).  The papers [ \!\!\citenum{Umbral}] specify the mock modular form by giving a recipe in terms of the root system of the Umbral
group to specify the shadow; imposing the existence of a pole of order $q^{-{1\over 4m}}$ at $\tau = i \infty$, with regular behavior at other cusps for
the twined functions; and specifying a slow-growth condition on the coefficients of the mock modular forms.

\medskip
There is also an associated umbral module, furnishing a statespace with symmetry $G$ and with an additional
${\bf Z}_{2m}$ quantum number.  The states at each energy level are governed by the umbral mock modular forms.  These modules have been proven to exist [\!\!\citenum{Gannontwo,DOG}], but have not been constructed in a physically convincing way yet.  There should hopefully be a uniform and physically transparent construction of all 23.

\medskip
The example to keep in mind is $A_1^{24}$.  The mock modular form in that case is $H(\tau)$ (in keeping with the discussion above, there is a 4 component vector-valued form -- but two components end up being 0, and the others are $\pm H(\tau)$).  A module with degeneracies governed by $H(\tau)$ was proven to exist by Gannon [\!\!\citenum{Gannontwo}].

\medskip
A secret underlying thought is that all of the umbral groups are related to symmetries of $K3$, and the different
umbral modules may somehow control transformations of the BPS states under symmetries originating from different
umbral groups\footnote{Here and in the following, by BPS states I simply mean states that preserve some fraction of the supersymmetry.  These states have usually played an important role in finding and testing dualities.}.  
A nice discussion of related issues, with a precise description of the appearance of all of the Umbral mock modular forms in the K3 elliptic genus, was given in [\!\!\citenum{MirandaSarah}].
There has been some thought of making this precise using Nikulin's theory of Niemeier markings [\!\!\citenum{Nikulin}] (for discussions, see e.g. [\!\!\citenum{Surfing,four}]).
Detailed studies gluing together symmetries of distinct K3 CFTs to try and understand the appearance of $M_{24}$ in the elliptic genus appear in [\!\!\citenum{Wendtwo,Wendthree}].
This can be considered a work in progress.

\subsection{Umbral groups and 3D string theory}

A physically clear relationship between Niemeier lattices, umbral groups, and string theory on K3 arises as follows [\!\!\citenum{us3D}].

\medskip
As you've likely heard, there are several different types of supersymmetric string theory.  Heterotic strings begin life in 10D with
gauge groups ($SO(32)$ or $E_8 \times E_8$) and half maximal supersymmetry.  Type IIA and type IIB string theory begin life with less gauge symmetry but more supersymmetry.
These theories are related by dualities in lower dimensions; for instance the heterotic string on $T^4$ is equivalent
to type IIA on $K3$, with (non-abelian) gauge groups on one side manifested in A-D-E singularities of the other side.  String duality is a large subject.  A nice general introduction is [\!\!\citenum{Vafaduality}], while a detailed discussion with more focus on K3 surfaces in string duality appears in [\!\!\citenum{Aspinwallreview}].

Let us now go on an apparent digression.  Consider the heterotic string compactification to 3D on $T^7$.
The moduli space of such models was discussed by Sen in 1994 [\!\!\citenum{Sen3D}] (building on
earlier work of [\!\!\citenum{Schwarz}]).  It is
$${\cal M}_{3D} = O(24,8;{\bf Z}) \backslash O(24,8) \slash (O(24)\times O(8))~.$$

This is interesting.  It is 192 dimensional.  Naively, the moduli space would be 162 dimensional:

\medskip
\noindent
* One gets 112 scalars in the 3D low energy theory from $E_8 \times E_8$ Wilson lines.

\medskip
\noindent
* The internal components of the flat metric on $T^7$ show up as Kaluza-Klein scalars in 3D.  There are 28 metric components on $T^7$.

\medskip
\noindent
* The heterotic string also has an antisymmetric two-form tensor field, $B_{\mu\nu}$.  There are 21 B-field components on $T^7$ which give rise to scalars in 3D.

\medskip
\noindent
* Finally, as with any string theory, there is the scalar which controls the string coupling -- the dilaton.

\medskip
However, in 3D one has a duality also between (abelian) gauge fields and scalars.  At generic points in
moduli space, the gauge group is $U(1)^{30}$ (16 factors from the $E_8 \times E_8$ or $SO(32)$ on its
Coulomb branch;  seven KK gauge fields from the metric; and 7 KK gauge fields from the B-field).  Dualizing
$$f_{\mu\nu}^{(a)} = \epsilon_{\mu\nu\rho}\partial_{\rho}\phi^{(a)}$$
gives us 30 additional scalars $\phi^{(a)}$.  The result is the 192 dimensional double coset moduli space
above.

\medskip
The duality group is larger than one would guess. 
Naively in $T^7$ compactification of heterotic strings, you'd get $O(23,7;{\bf Z})$.  The additional elements
enlarging this to $O(24,8;{\bf Z})$ morally come from {\bf S-dualities} visible already in four dimensions.

\medskip
Why are we taking this detour to three dimensions?  Because in the 3D string theory, Niemeier lattices and
their umbral symmetry groups play a preferred role in the physics [\!\!\citenum{us3D}].  Notice that ${\cal M}_{3D}$ can be viewed
as the moduli space of even unimodular lattices of signature (24,8).  There are 24 special points in the moduli
space where the lattice splits as
$$\Gamma^{24,8} ~=~\Gamma^{24,0} \oplus \Gamma^{0,8}~.$$
$\Gamma^8$ must be the $E_8$ root lattice, but $\Gamma^{24,0}$ can be the Leech lattice or any of the 23
Niemeier lattices.

\medskip
Note that all of these ``Niemeier points" in moduli space are non-perturbative phenomena.  They cannot be
seen perturbatively in any duality frame in 3D. 

\medskip
The theory at the Niemeier points has enhanced $A-D-E$ gauge symmetry.  A small perturbation breaks the 
symmetry around any such point to $U(1)^{30}$ while preserving the umbral symmetry group $G$ associated to
the relevant Niemeier lattice.  The umbral groups can be seen perturbatively in the heterotic theory both in 3D and (even more obviously)
in 2D.

\medskip
This gives a good way to think of the umbral symmetries in $K3$.  Heterotic string theory on $T^7$ is dual to type II string theory on 
$K3 \times T^3$.  The picture you should keep in mind is shown in Figure 3.

\begin{figure}
\label{umbral}
\begin{center}
\includegraphics[width=0.55\textwidth]{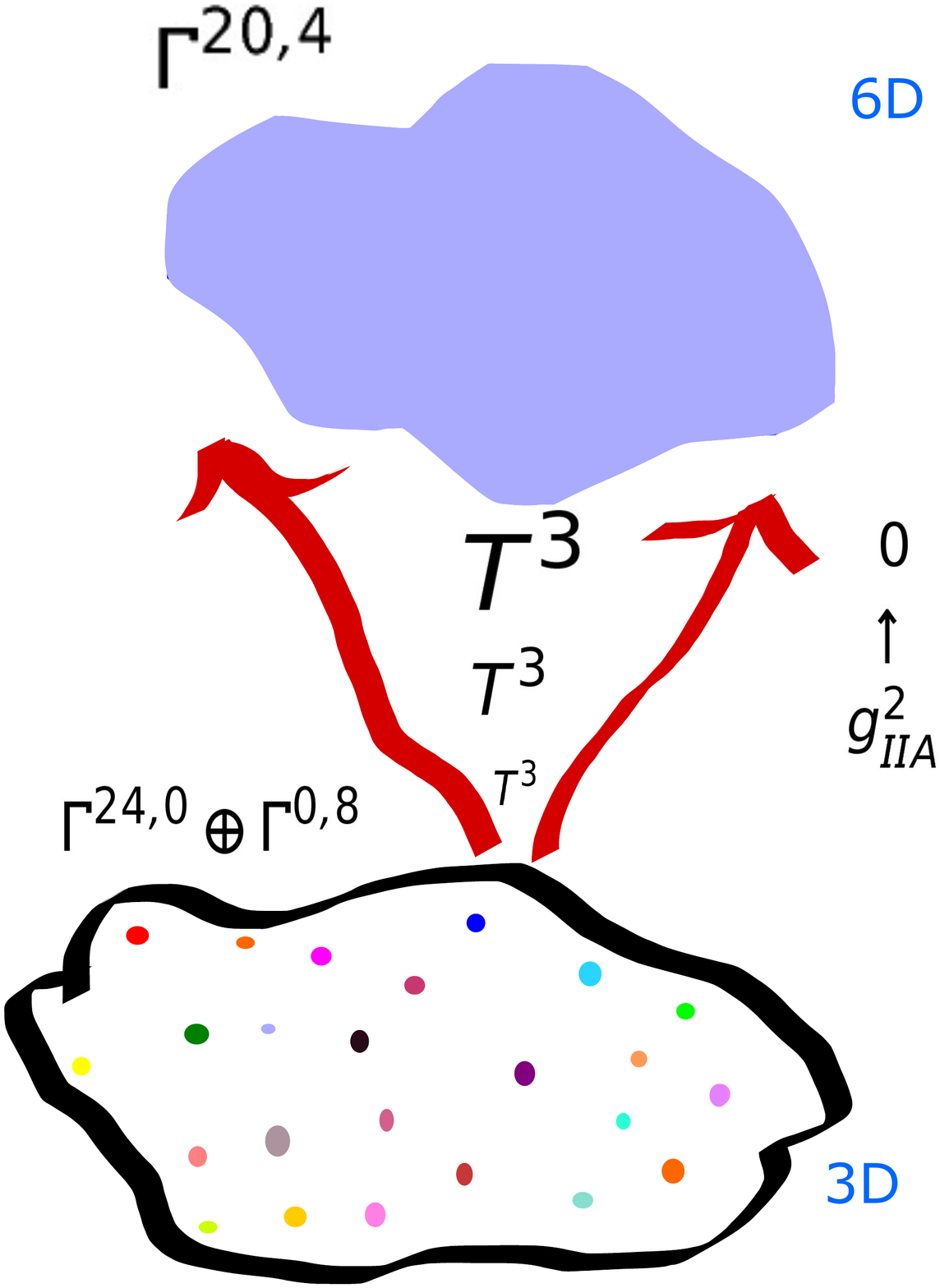}
\end{center}
\caption{How to associate 6D symmetries with Niemeier points.  Figure from [\!\!\citenum{us3D}].}
\end{figure}

\medskip
\noindent
 One can obtain 6D theories on K3 by decompactifying a $T^3$.  This requires a choice of a
4-plane in the $\Gamma^{24,8}$.  This is consistent with the findings (from many perspectives)
that symmetries of K3 sigma models correspond to {\bf 4-plane} preserving subgroups of 
$Co_0$.  

\medskip
\noindent
Symmetries of the 6D theory which can be followed continuously from 3D starting from the vicinity of
a Niemeier point, should be associated with that Umbral group and have twinings governed by the 
relevant Umbral moonshine.

\subsection{1/2 BPS computations}

We would now like to read off the spectrum and symmetry properties of the BPS states.
The basic strategy we will pursue is the following.  After string duality, it is most natural to focus on physical quantities which appear in the space-time effective action -- as these are duality invariant.  So to discuss a BPS spectrum, it is best to find a coupling function which is governed by BPS states, and constrain its form.
In theories with 8 supercharges, famously, functions like the gauge coupling function appearing in front of the
Maxwell terms in the Lagrangian receive interesting quantum corrections [\!\!\citenum{SW}].  In theories like those under
consideration here, with sixteen supercharges, the situation is more constrained.  The two-derivative effective Lagrangian does not have interesting corrections.  However, there are 
{\bf four-derivative} terms that do receive interesting and computable corrections.
Very roughly, these are terms of the form
$$f(\phi) (\partial\phi)^4 \subset {\cal L}_{3D}~.$$
and we are computing the function $f(\phi)$ for some very special choices of the contractions of indices
labelling the moduli. (The full story of these couplings can be found in the work of Obers and Pioline,
see [\!\!\citenum{Obers}]).

Some fully explicit computations of these four-derivative terms in ${\cal L}_{3D}$, which are saturated by 1/2 BPS states, can
be found in [\!\!\citenum{us3D}].  These exhibit e.g. a decomposition of the 1/2 BPS states 
into $M_{24}$ irreps near the $A_{1}^{24}$ point in moduli space.

Here, we just give a flavor of the results.  Specialize to a point in moduli space where 
$$\Gamma^{24,8} \sim L \oplus E_8(-1)$$ 
with $L$ a positive definite, even unimodular 24 lattice.\footnote{Here, the notation $E_8(-1)$ simply means that
we consider the $E_8$ lattice with the normal quadratic form measuring norms flipped in sign.}  For any $\lambda \in L$, we can
consider a theta series
$$\Theta_{L,\lambda}(\tau,z) = \sum_{\lambda^\prime \in L} q^{(\lambda^\prime)^2 \over 2} y^{\lambda^\prime \cdot
\lambda} ~=~\sum_{n,r \in {\bf Z}} c_{L,\lambda}(n,r) q^n y^r~.$$
This series will have cofficients governing the four-derivative correction 
as we move in moduli space along a direction determined (infinitesimally) by the choice of $\lambda$.
You can understand this because the $24$ vector multiplets of the 3d $N=8$ supersymmetry each
have scalars in them (8 after dualizing, to be precise), and the choice of $\lambda$ is choosing a direction in the 24 dimensional
space of vector multiplets.  (Which of the 8 scalar components in a given scalar in the 3D vectormultiplet that we pick, doesn't matter much).

It follows from standard facts about lattice theta functions that this will be a holomorphic Jacobi form of weight
$$w = {1 \over 2} {\rm dim}(L)$$
and index
$$m = {\lambda^2 \over 2}~.$$
The Fourier coefficients are
$$c_{L,\lambda}(n,r) = \# \{ ~\lambda^\prime \in L~ |~ (\lambda^\prime)^2 = 2n,~\lambda \cdot \lambda^\prime = r~\}~$$

The choice of 
$\lambda$ breaks the group $Aut(L)$ down to $H_{\lambda} \subset Aut(L)$ which fixes $L$.  
Recall that for a given instance of umbral moonshine, $Aut(L) = W \rtimes G$ with $W$ the relevant
Weyl group and $G$ the umbral group.  There exist choices of $\lambda$ (proportional to a Weyl vector)
that result in $H_{\lambda} \simeq G$.

For the Niemeier lattice $A_{1}^{24}$ with $\lambda$ equal to a Weyl vector, we find
$\Theta_{L,\lambda}$ to be of weight 12 and index 6.  One can present a fully explicit expression for it in terms of
basis Jacobi forms [\!\!\citenum{us3D}].  The $M_{24}$ representations appearing in low lying terms in the (q,y) expansion are
$$c(1,1) = 24 = {\bf 1} + {\bf 23}$$
$$c(2,4) = 759 = {\bf 1} + {\bf 23} + {\bf 252} + {\bf 483}$$
$$c(2,3) = 6072 = {\bf 1} + 2 \times {\bf 23} + 2 \times {\bf 252} + {\bf 253} + {\bf 483} + 
{\bf 1265} + {\bf 3520}$$
$$ \cdots $$

\medskip
It remains a very interesting challenge to find a physical role for the umbral mock modular forms in the 3D picture.  It is likely that this will come from a deeper understanding of 1/4 BPS states.  It would also be
very interesting to give a precise connection between the association of umbral symmetries with K3 surfaces following from this picture, and the theory of Niemeier markings.  Finally, it is important to mention that while my own focus on manifesting the symmetries of the problem has led me to emphasize a possible role for
3D string vacua in explaining umbral moonshine, from other perspectives, other limits of the theory could play roles of equal or greater importance in our eventual understanding.

\section{Lecture 4: Moonshine and AdS3 quantum gravity?}

In the next two lectures, we move on to some connections of the ideas we've developed to other areas of
theoretical physics.  Here, we focus on some (possibly coincidental) connections between the CFTs
involved in moonshine, and theories that could be related to AdS3 quantum gravity.  We begin with a number of
disclaimers:

\medskip
\noindent
* There are various good reasons to be skeptical that weakly curved AdS3 quantum gravity (with no matter
fields) exists.  See the discussion in [\!\!\citenum{MaloneyWitten}].

\medskip
\noindent
* There are reasons to think that perhaps the conditions espoused here to define ``pure quantum gravity," could
or should be modified.  See e.g. the very recent [\!\!\citenum{Liam}] for one proposed modification.

\medskip
We will blithely ignore these complications and proceed.  The ideas are interesting enough to justify hearing about them once.

\subsection{Criteria on pure AdS3 gravity}

The AdS/CFT correspondence was famously derived by taking a near-horizon limit, using ``real D-branes" in string
theory [\!\!\citenum{Maldacena}].  But we can abstract the basic facts of the duality, without a particular D-brane construction in mind.

\medskip
One of the most basic elements of the map takes the local (scalar) operators in the CFT to local bulk fields in 
AdS.  The map for $AdS_{d+1} \leftrightarrow CFT_d$ takes a local scalar operator of dimension $\Delta$,
to a bulk field whose mass scales as 
$$m^2 ~\sim~\Delta(\Delta - d)~.$$

\medskip
There are many theories we might wish to study, that we do not (yet) know how to find via scaling limits of D-branes.
But this doesn't mean we can't study their possible AdS/CFT dualities.  Let us pick the simplest: {\bf pure} gravity
in AdS3.  The main ingredient we will use is the map between bulk fields and boundary operators, and what this tells us about the spectrum of a CFT dual to pure gravity.

\subsubsection{The first attempt}

By pure gravity, we mean the theory with no bulk fields other than the multiplet dual to the stress tensor.
We will therefore postulate a CFT whose Hilbert space contains nothing but the vacuum and its descendants.
Such a hypothetical CFT won't be consistent, but lets see how far we can get.  Our discussion closely follows the ideas of
[\!\!\citenum{Witten3D}].

\medskip
A universal fact about AdS3/CFT2 was discovered by Brown and Henneaux in 1986 [\!\!\citenum{Brown}].  It is that 
$$c = 3 {L_{AdS} \over 2G}$$
where $c$ will be the central charge of the dual 2d CFT, $L_{AdS}$ is the curvature radius of AdS3, and $G$ is the Newton constant.  It is important to stress that Brown and Henneaux comptued this $c$ not by knowing a dual 2d CFT, but by thinking about commutators of symmetry generators visible purely in the gravitational
theory.
The Virasoro symmetry shows up via symmetries visible at spatial infinity in the AdS3 gravity theory, and $c$ can be computed a priori from 
commutators in the gravity side.  Importantly, c obtained in this way {\bf does} match the central charge of the dual CFT in existing dual pairs.

\medskip
Now, a standard count of the number of physical polarizations after removing gauge redundancies 
reveals that there are no propagating gravitons in 3D gravity.  But there are AdS ``boundary gravitons." These correspond to 
the descendants of the AdS vacuum (in the dual CFT), with respect to the Virasoro generators.  I.e., we can consider
strings of Virasoro operators acting on the $SL(2,{\bf R})$ invariant vacuum 
$|0\rangle$,
$$L_{-n_1}L_{-n_2} \cdots L_{-n_k}|0\rangle~.$$

\medskip
The counting of the states will be as follows.  $L_{-1}$ annihilates the vacuum.  But we could act with $L_{-2}$ any number of times:
$$|0\rangle,~L_{-2}|0\rangle, ~L_{-2}^2 |0\rangle, \cdots$$
This gives a tower of states that would contribute, to the torus partition function of a would-be dual CFT,
$${1\over (1-q^2)}~.$$
Similarly, the $L_{-n}$ (with $n > 2$) would yield a factor of
$${1\over (1-q^n)}~$$
in the torus partition function.
So we find for the full contribution of the descendants,
$$\prod_{n\geq 2}{1\over (1-q^n)}~.$$

\medskip
This isn't quite right yet, as we need to remember that the vacuum contribution is weighted with a $q^{-{c\over 24}}$ in the
partition sum, because of the Casimir energy.  For $c=24k$, this would give
$$Z(q) = q^{-k} \prod_{n \geq 2} {1 \over (1-q^n)}~.$$
This is the would-be partition function of the CFT dual to pure quantum gravity in AdS3, with cosmological
constant $\sim {1\over k}$.

\medskip
A few comments:

\medskip
\noindent
$\bullet$ The astute reader will note that I have assumed that the CFT is chiral.  There could be various justifications
for this.  One is that we could be looking for duals to `chiral gravity' [\!\!\citenum{chiralgravity}].  Another is that there are indications from
the Chern-Simons formulation of 3d gravity that the values $c=24k$ are preferred, which allows for (but does not
guarantee!) holomorphic factorization [\!\!\citenum{Witten3D}].  In any case, I am going to continue with this assumption for now. 

\medskip
\noindent
$\bullet$ This $Z(q)$ is obviously not the torus partition function of any CFT.  It is not modular invariant.

\medskip
Can we fix it?

\subsubsection{Fixing it}

Let us think a bit about the physics. We will be sloppy with O(1) factors, simply trying to give the ``big picture."  The basic spectrum of AdS gravity is quite
simple.  There are particles with mass $< c$ (in units of the AdS radius).  These are perturbative particles in AdS,
quanta of bulk quantum fields.

\medskip
There are also objects with mass $\geq c$.  These are {\bf black holes} in AdS, the BTZ black holes.  

\medskip
While we know a lot about the spectrum of perturbative particles in a theory of gravity coupled to matter, we can be a bit more agnostic about our knowledge of the black holes.  

\medskip
Where do these things appear in the partition function?  After restoring factors of O(1), it turns out that the terms
$$q^{-k} + c_1 q^{-k+1} + \cdots + c_{k}q^0$$
in $Z(q)$ are the ones corresponding to perturbative particles.  In a theory of pure gravity, we could reasonably
require that the coefficients $c_i, i \leq k$ be determined by the Virasoro descendants of $|0\rangle$.

\medskip
What about the terms
$$c_{k+1} q + c_{k+2} q^2 + \cdots ~?$$

\medskip
Now, we use some facts about modularity.  The (chiral) partition function is a modular function.
On physical grounds, we expect poles at $\tau = i \infty$ but not at interior points in moduli space.
So at $c = 24k$, the partition function is a polynomial in $J(q)$ of degree $k$
$$Z = J^k + d_1 J^{k-1} + \cdots$$
There are precisely $k$ undetermined coefficients here.  And they can be precisely determined by matching
the polar terms (including the term $q^0$) in the partition function with only descendants of $|0\rangle$
up to $O(q^0)$.

\medskip
This means that, with the assumption that we transition from perturbative particles to black holes at 
energy $k+1$ above the vacuum in the theory with $c = 24k$, there is a uniquely predicted partition function
that is modular invariant.  The possible existence of ``extremal" CFTs with these partition functions was
first discussed by H\"ohn [\!\!\citenum{Hohn}].

\medskip
\noindent
Some comments:

\medskip
\noindent
$\bullet$ There is no real justification for assuming that the transition from known descendants to ``unknown" and hence not precisely predicted black holes, occurs at $q^0$.  It could as well occur at any $q^{\alpha}$ with
$$|\alpha| \ll c$$
as $c \to \infty$.  But this would be less restrictive, and so we may as well pursue the most constrained story
to its logical conclusion first.

\medskip
\noindent
$\bullet$ Just because we can predict a partition function, does not mean that a CFT with this partition function
exists (or, that it is unique if it does exist).  This is the question we will now explore.

\subsection{Examples}

\subsubsection{The Monster}

So, can we construct theories with the desired partition functions?  Lets start at the simplest place.  As
$c = 24k$, this is at $c=24$.  The partition function that matches the q-expansion of descendants of $|0\rangle$,
through the $q^0$ term, is
$$Z = J(q) = q^{-1} + 196884 q + \cdots~.$$

\medskip
Happily, we already know of a theory with this partition function!  We encountered it in Lecture 1.  It is precisely the theory arising in
Monstrous moonshine, the Leech CFT mod ${\bf Z}_2$.

\medskip
It would certainly be intriguing if there was a connection between moonshine and quantum gravity in AdS.
One place where a tie to 3D quantum gravity could help shed light on Monstrous moonshine, is in providing a 
possible conceptual explanation for the Hauptmodul property discussed in Lecture 1 \cite{Igor}.

\medskip
Here is one amusing extension of this observation.  At the $q^1$ term, we have our first ``black holes."
The Bekenstein-Hawking entropy of a state at a given
$L_0$ eigenvalue (and at central charge $24k$) is
$$S_{BH} = 4\pi \sqrt{kL_0}~.$$
But, we have the exact degeneracy of the first excited states in the Monster CFT,
$${\rm log}(196883) \simeq 12.19~.$$
Meanwhile,
$$4\pi \sqrt{1} \simeq 12.57~.$$
Not bad!

Sadly, theories exhibiting the $k=2,3,\cdots$ partition functions have eluded construction so far. 
There have been some discussions of properties such theories would have [\!\!\citenum{GaiottoYin,Yin,Gaiotto,GKV}]
and some suggestive arguments
against existence [\!\!\citenum{Gaberdiel}], but there is as yet no proof or disproof that such
theories exist.

\subsubsection{Adding supersymmetry}

To get more examples, we can add supersymmetry.  With
$N=1$ supersymmetry, some important things in our 
discussion change.

\medskip
\noindent
* It is now natural to consider partition functions
which are invariant under a smaller modular group
than $SL(2,{\bf Z})$ -- namely, $\Gamma_0(2)$ or other
index 3 subgroups.  Fermions, after all, have half-integer
modes when they have anti-periodic (NS) boundary conditions.

\medskip
\noindent
* We can repeat the same logic with the polynomials generating
potential partition functions for ``pure N=1 supergravity."  We try to match a hypothetical
partition function that counts only the vacuum state and (super)Virasoro descendants, up to the
$q^0$ term.
What will change is that now $c = 12k$, and there is a new 
Hauptmodul, commonly called $K$ instead of $J$.  It is given
by
$$K(\tau)  = {\eta(\tau)^{48} \over {\eta(2\tau)^{24}\eta({\tau \over 2})^{24}}} -24~.$$

\medskip
\noindent
* And again, at $k=1$, there is a theory!  This is again a famous
theory in the context of moonshine.

\medskip
There are two very simple chiral $c=12$ CFTs that might occur to one.
One is based on the super-E8 lattice.  (After all, the E8 lattice is
even and unimodular).  Another is based on 24 chiral free fermions.

\medskip
Neither has the ``right" partition function for k=1 supergravity, but
the ${\bf Z}_2$ orbifolds of both (where the ${\bf Z}_2$ acts by inverting all fields) give the same CFT, which does!

$$Z_{NS}(\tau) = {\rm tr}_{NS} q^{L_0 - c/24} =
{1\over 2}\left({E_4 \theta_3^4 \over \eta^{12}}
+ 16{\theta_4^4 \over \theta_2^4} + 16 {\theta_2^4 \over
\theta_4^4}\right) (\tau)~=~K(\tau)!$$
This gives a q-expansion
$$K(\tau) = q^{-1/2} + 0 + 276 q^{1/2} + 2048 q + 11202 q^{3/2} + \cdots~.$$

\medskip
It is not a coincidence that the $Co_0$ group -- a double cover of Conway's largest sporadic group -- 
has representations that match these coefficients:
$$2048 = {\bf 24} + {\bf 2024}$$
$$11202 = {\bf 1} + {\bf 276} + {\bf 299} + {\bf 1771} + {\bf 8855}$$
$$\cdots$$

\medskip
As with Monstrous moonshine, the existence of these decompositions is not coincidental.  There
is a moonshine relating Conway's largest sporadic group to this 2d CFT.
This theory was briefly discussed by Frenkel-Lepowsky-Meurman, and the full story of the $Co_0$ 
moonshine has been described by Duncan [\!\!\citenum{Duncan, DuncanMC}].  

\medskip
This particular theory has also played other interesting roles in moonshine -- in the first rigorous
examples of mock modular moonshine involving various sporadic simple groups [\!\!\citenum{six}], and in connection with modules governing refined BPS invariants
of K3 [\!\!\citenum{K3inv}].  The latter will be the subject of Lecture 5.

\medskip
Once again, there is no proof that $N=1$ extremal theories do or do not exist at large $k$.
At $k=2$, however, such a theory does exist.  One can equip the Monster CFT with a hidden $N=1$
supersymmetry [\!\!\citenum{DGH}], and viewed as an $N=1$ theory, it is again extremal!

\subsubsection{Adding more supersymmetry}

If we move up to $N=2$ supersymmetry, we can make further progress.  We can prove that with the minimal 
assumption, extremal $N=2$ SCFTs will not exist at large central charge.  I will summarize the simple logic, following work of
Gaberdiel-Gukov-Keller-Moore-Ooguri [\!\!\citenum{GGKMO}].

\medskip
An $N=2$ SCFT with $c = 6m$ has an elliptic genus, as we discussed starting in Lecture 2:
$$Z_{EG} = {\rm Tr}\left( q^{L_0-{c\over 24}} y^{J_0} \bar q^{\bar L_0-{c\over 24}} (-1)^{F_L+F_R}\right)~.$$
This is a weak Jacobi form - and nice and holomorphic - {\bf without} making any assumption analogous
to holomorphic factorization.  That is a good thing (though the examples we will find later of extremal $N=2$ theories at
small central charge
will happen to be chiral theories in any case).
It has weight 0 and index $m$.  We will focus on the case of integer $m$ here.

\medskip
As we discussed in Lecture 2, a basis for such forms is given by $E_4, E_6$ together with
$$\phi_{-2,1}(\tau,z) = {\theta_1(\tau,z)^2 \over \eta^6}$$
$$\phi_{0,1}(\tau,z) = 4 \left( {\theta_2(\tau,z)^2 \over \theta_2(\tau,0)^2} + {\theta_3(\tau,z)^2 \over \theta_3(\tau,0)^2} + {\theta_4(\tau,z)^2 \over \theta_4(\tau,0)^2}\right)~.$$

\medskip
Now, we need to worry about the {\bf polar terms} in the elliptic genus (we will give a precise definition shortly).  The strictly analogous assumption to the one
we were making before, is that these polar pieces should match those coming from the vacuum character of the
$N=2$ superconformal algebra.  

\medskip
Recall that the general expansion of the elliptic genus looks like
$$Z(\tau,z) = \sum_{n \geq 0,l} c(n,l) q^n y^l~.$$
Based on the spectral flow automorphism of the $N=2$ algebra [\!\!\citenum{Schwimmer}]
$$L_0 \to L_0 + \theta J_0 + \theta^2 m$$
$$J_0 \to J_0 + 2\theta m~,$$
the terms with $|l|\leq m$ are sufficient
to determine the genus. (An additional symmetry relating $c(n,l)$ to $c(n,-l)$ further reduces the independent coefficients).   
See Figure 4 to understand the geometry of the independent polar terms.

\begin{figure}
\label{sporadic}
\begin{center}
\includegraphics[width=0.75\textwidth]{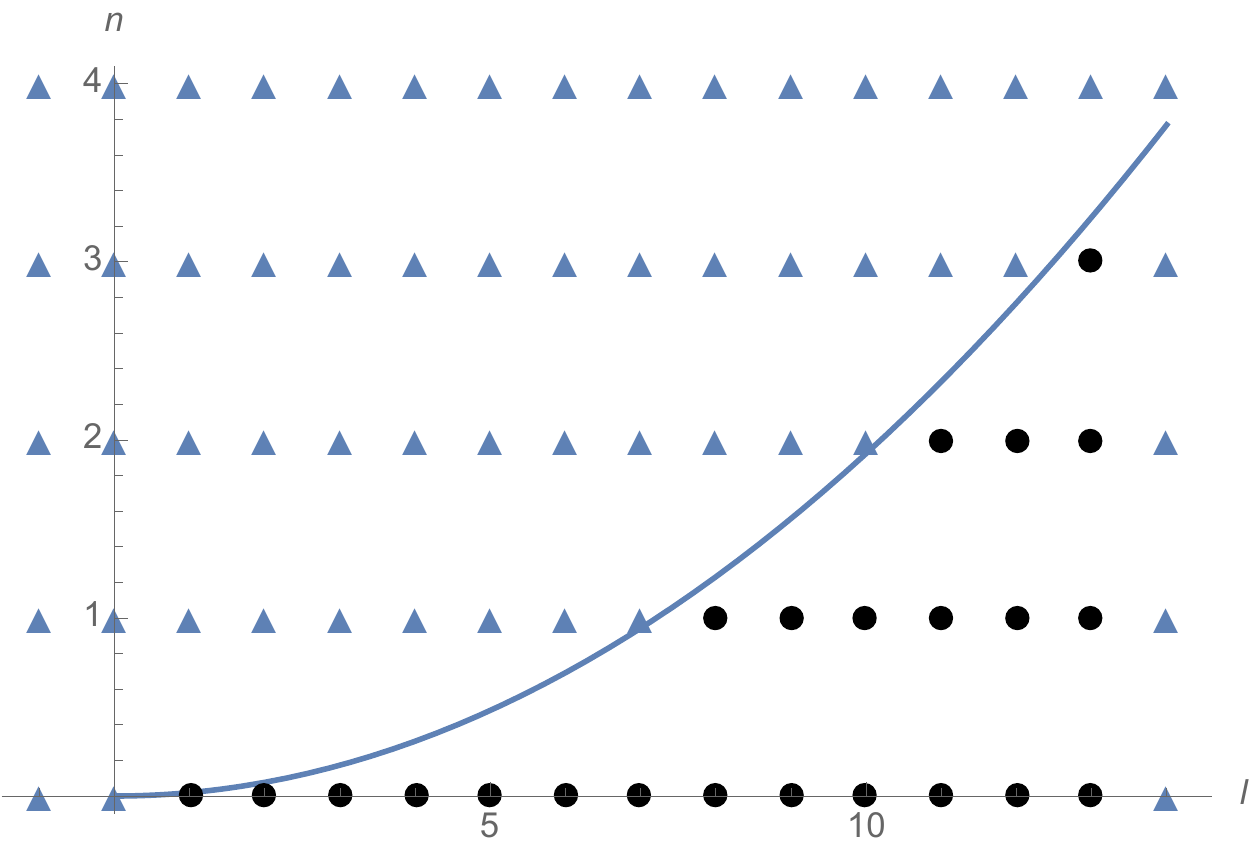}
\end{center}
\caption{Plot of the polar terms in (n,l) space. The curve is $n = l^2/4m$ for $m=13$.  The circles mark the integral points in the polar region, while the triangles
do not give independent polar states.}
\end{figure}

\medskip
The {\bf polarity} is defined as
$$p = 4mn - l^2~.$$
It is called polarity because the terms with $p<0$ correspond to polar coefficients in the ordinary modular forms governing the theta series of a Jacobi form.
In the naive semiclassical map to gravity, the terms with $p > 0$ are black holes, while terms with $p \leq 0$ correspond to supergravity states.
The polarity is bounded below by $-m^2$, so this leaves a finite number of possible polar terms.  For more discussion of the relationship with states at a positive polarity to black holes, see e.g. [\!\!\citenum{Larsen,Farey,elliptic}].

\medskip
Now, finally, with the power of $N=2$ supersymmetry, we will be able to state an interesting general result.

\medskip
\noindent
$\bullet$ The number of polar coefficients, is
$$P(m) = {m^2 \over 12} + {5m \over 8} + O(m^{1/2})$$
at large $m$.  This is easily obtained by doing a precise count of integer points contained
under the curve in Figure 4.

\medskip
\noindent
$\bullet$ The dimension of the space of weak
Jacobi forms at a given (large) $m$ is
$$j(m) = {m^2 \over 12} + {m\over 2} + \cdots$$
This can be obtained by using the explicit basis forms we described and enumerating.

\medskip
\noindent
So there is a simple and suggestive computation.
$$\# ({\rm polar~coefs}) - {\rm dim(space ~of ~forms)} = P(m) - j(m) = {m \over 8} + \cdots$$

\medskip
Under the (not unreasonable) assumption that the polar coefficients induced from the
vacuum character are somewhat typical, and do not lie in the lower-dimensional vector
space characterizing true weak Jacobi forms ``by accident," we then expect that at sufficiently 
large $m$, $N=2$ extremal theories do not exist!

\medskip
More rigorous reasoning suffices to verify this [\!\!\citenum{GGKMO}].  Again, the important caveat that one
can modify the condition for polar coefficients at energies of $O(c)$ but before the 
naive onset of `black holes,' needs to be emphasized.  This could change the answer to 
the question of whether such theories exist.

\medskip
Again, at low $c$, such theories {\bf do} exist.  And again, they are relevant to moonshine!
See [\!\!\citenum{six,usextremal}] for discussions of pure $N=2$ theories at 
$c=12, 24$ which enjoy a connection with moonshine for Mathieu groups.  

\medskip
We could
also continue the discussion to further extended supersymmetry, and find examples at
low central charge.  We will not pursue this here, but see for instance
[\!\!\citenum{Sarah}] for discussion of an example of an $N=4$ extremal theory.\footnote{In fact, 
we are informed by Sarah Harrison that the $c=12$ theory studied in [\!\!\citenum{spinseven}] is also
extremal with respect to its extended (super) W-algebra.}

\medskip
In closing this section, we can summarize by saying that while several examples of extremal CFTs with a variety
of (super)algebras have been discovered, all have $c \leq 24$.  While construction methods of CFTs with sparse spectrum are primitive, and so we may 
simply be missing examples which exist with the stated criteria in some cases, it is also a logical possibility that (slightly) relaxing the
criteria at $c > 24$ (but maintaining a paucity of low-energy particles) would lead to further progress.

\section{Lecture 5: Enumerative geometry and moonshine}

In this final section, we close with one more ``application" of moonshine, this time in mathematics.  We will describe some connections between a moonshine module -- appearing
in the Conway moonshine discussed in Lecture 4 -- and the enumerative geometry of K3 surfaces.
We will see that the hierarchy of K3 invariants given by the Yau-Zaslow nodal curve counts [\!\!\citenum{YZ}], the
KKV invariants [\!\!\citenum{KKV}] including one further grading, and the KKP invariants [\!\!\citenum{KKP}] including
two further gradings, are all natural traces in this moonshine module.  This is in accord with the 
fact -- discussed in Lectures 2,3 -- that all symmetries of K3 CFTs can naturally be embedded in 
$Co_0$.

\subsection{Counting BPS states on K3}

Counts of BPS states in string theory have been useful in obtaining an understanding of black hole entropy; in elucidating the behavior of soluble field theories which arise in decoupling limits of string compactification; and in providing duality-invariant ``tags" of the theory, providing highly nontrivial consistency checks on various dualities.  There is some hope that proper understanding of the BPS states may also lead to a deeper understanding of the mathematics underlying string theory.

In the simple setting of heterotic or type IIA compactifications (on $T^4$ or $K3$, respectively) with 16 supercharges, the 1/2 BPS states can be counted following the logic of Dabholkar and Harvey [\!\!\citenum{Dabholkar}].  The heterotic string has all supersymmetries coming from the right movers, while the left-moving degrees of freedom are those of a
$c=24$ bosonic string.  To find BPS states, we therefore leave right-movers in a ground state, but freely excite the left-moving oscillator modes.  The result is a 1/2 BPS counting function

$$N_{BPS}(q) = \sum_{n \geq 0} d_n q^{n-1} = {1\over \Delta(\tau)} = {1\over \eta^{24}(\tau)} = 
q^{-1} ( 1 + 24 q + 324 q^2 + 3200 q^3 + \cdots)$$

\noindent
with $d_n$ governing the number of BPS states at a given mass level.

We can turn this into a more geometric statement in two ways.  In the type IIA string on K3, which is dual
to heterotic strings on $T^4$, the BPS states can be counted in one duality frame by studying bound states
of $n$ D0-branes to a D4-brane [\!\!\citenum{Vafacount}].  As the moduli space of a D0-brane on K3 is a copy of the
space itself, this thinking leads to a count
$$N_{BPS}(q) = \sum_{n \geq 0} \chi(K3^{[n]}) q^{n-1}~.$$
Here $K3^{[n]}$ denotes the Hilbert scheme of n points on $K3$; it is a desingularization of the $n$th
symmetric product of a K3 surface.  In fact, G\"ottsche has shown [\!\!\citenum{Gottsche}]
$${1\over \Delta(\tau)} = \sum_{n \geq 0} \chi(K3^{[n]}) q^{n-1}~,$$
and so this alternate count of BPS states agrees with the heterotic string, as expected from string duality.

An alternative geometrization arises in yet another duality frame.  Instead of counting 
$n$ D0-branes in a 
D4-brane, we can get a BPS state with the same charges by considering a D2-brane wrapping a 
curve with self-intersection number $2n-2$.  This can be realized on a holomorphic curve of genus $n$
in K3.  D2-branes come equipped with a $U(1)$ gauge field, so the count is now governed by the
Euler character of the moduli space ${\cal M}_{n}^H$ of holomorphic curves of genus $n$ with a choice of flat
$U(1)$ bundle [\!\!\citenum{BSV}].  It was argued by Yau and Zaslow [\!\!\citenum{YZ}] and proved in
[\!\!\citenum{NLtheory}] that the contribution to $\chi({\cal M}_{n}^H)$ localizes on curves of genus 0 with $n$ double points.  Hence we get a Yau-Zaslow formula
$$N_{BPS}(q) = \sum_{n \geq 0} {\#}_{\rm nodal}(n) q^{n-1} = {1\over \Delta(q)}~.$$
In this way, we relate a counting function of interest in enumerative geometry, to a BPS state count
governed by G\"ottsche's formula, or by the heterotic string.

\subsection{Refining the counts}

Of course, there is more to the life of a BPS state than simply its mass.  For instance, in string compactification
to 5d on $K3 \times S^1$, there is an $SO(4) \simeq SU(2)_L \times SU(2)_R$ little group in the non-compact dimensions.  So we can
imagine grading the count of BPS states by the $SU(2)$ quantum numbers.

Let us work our way up in steps.  Returning to G\"ottsche's formula, we can clearly further refine the
Euler character to the $\chi_y$ genus:
$$\chi_y(M) = \sum_{p,q} y^p (-1)^{q} h^{p,q}(M)$$
with $h^{p,q}$ the Hodge numbers.  Then it is natural to write the generating function
$$\sum_{n \geq 0} \chi_{-y}(K3^{[n]}) y^{-n} q^{n-1} = q^{-1} \prod_{k>0} (1-yq^k)^{-2} (1-q^k)^{-20}
(1-y^{-1}q^k)^{-2}$$
$$ = (-y + 2 - y^{-1}) {\eta(\tau)^{6} \over \theta_{1}(\tau,z)^2} {1\over \Delta(\tau)}~$$
with 
$$y = e^{2\pi i z}~.$$
And this formula, too, admits a curve-counting 
interpretation.

Define the numbers $n_{n}^{r}$ through the formula
$$\sum_{r \geq 0}\sum_{n\geq 0} (-1)^r n_n^r (y^{1/2} - y^{-1/2})^{2r} q^{n-1} = 
q^{-1} \prod_{k>0} (1-yq^k)^{-2}(1-q^{k})^{-20}(1-y^{-1}q^k)^{-2}~.$$
Then it was proposed by Katz-Klemm-Vafa that the numbers $n_{n}^r$ encode the reduced
Gromov-Witten invariants of a K3 surface.  
This reduces to the Yau-Zaslow counting formula through the specialization 
$$d_n = n_{n}^{r=0}~.$$
The KKV conjecture was proved in [\!\!\citenum{PT}].

Finally, there is a natural refinement to the full generating function for Hodge polynomials of
$K3^{[n]}$ [\!\!\citenum{Gottscheagain}].  Recall that
$$\chi_{\rm Hodge}(M) = u^{-d/2}y^{-d/2} \sum_{p,q} (-u)^{q}(-y)^{p} h^{p,q}(M)$$
for a K\"ahler manifold of complex dimension $d$.
Then the fully refined formula that we wish to interpret geometrically is
$$\sum_{n \geq 0} \chi_{\rm Hodge}(K3^{[n]}) q^{n-1}$$
$$= q^{-1}\prod_{k>0} (1-uy  q^k)^{-1} (1- u^{-1}y q^k)^{-1} (1-q^k)^{-20} (1- u y^{-1} q^k)^{-1} (1- u^{-1}y^{-1} q^k)^{-1}$$
$$= (u-y-y^{-1}+u^{-1}) {\eta(\tau)^6 \over \theta_1(\tau,z+w) \theta_1 (\tau,z-w)} {1\over \Delta(\tau)}~$$
where
$$u = e^{2\pi i w}~.$$

The geometric interpretation of the formula with two additional gradings is as follows.  
Define
$$[j]_x \equiv x^{-2j} + x^{-2j + 2} + \cdots + x^{2j}$$
for $j \in {1\over 2}{\bf Z}$.  Then
$$\sum_{n \geq 0} \sum_{j_L,j_R \in {1\over 2}{\bf Z}^+} N_{n}^{j_L,j_R} [j_L]_y [j_R]_u q^{n-1}$$
$$= q^{-1}\prod_{k>0} (1 - uyq^k)^{-1} (1-u^{-1}yq^k) (1-q^k)^{-20} (1-u y^{-1} q^k)^{-1}
(1-u^{-1}y^{-1}q^k)^{-1}~.$$
We define $N_{n}^{j_L,j_R}$ as follows.
$N_{\beta}^{j_L,j_R}$ is a refined Gopakumar-Vafa invariant. 
 Physically, it is a BPS index counting the
BPS states of M2-branes wrapping a 2-cycle in homology class $\beta$ with spin quantum numbers $(j_L,j_R)$ under the
$SU(2)_L \times SU(2)_R$ little group mentioned above.  The label $n$ arises through the conjecture of
[\!\!\citenum{KKP}] that for a K3 surface $X$, $N_{\beta}^{j_L,j_R}$ depends only on 
$$ \beta \cdot \beta = 2n -2 $$
for any $\beta \in H_2(X,{\bf Z})$.

We've now achieved an enumerative interpretation of the BPS counting function with gradings
corresponding to each of the $SU(2)$s.  The KKV formula can clearly be interpreted as the specialization keeping account of
a single $SU(2)$ grading.

\subsection{Conway moonshine and the K3 elliptic genus}

In Lecture 4, we encountered a simple chiral CFT which enjoys moonshine for Conway's largest sporadic group
$Co_1$ (or really its double-cover, $Co_0$).  This is a theory based on the ${\bf Z}_2$ orbifold of super-E8 current algebra.
It has an equivalent formulation as the ${\bf Z}_2$ orbifold of 24 free real chiral fermions $\psi_i$ by
$$\psi_i \to - \psi_i~.$$
The NS sector partition function is
$$Z_{NS}(\tau) = {\rm Tr}(q^{L_0 - {c\over 24}})$$
$$ = {1 \over 2}{1\over \eta^{12}(\tau)} \sum_{i=2}^{4} \theta_{i}^{12}(\tau,0) = q^{-1/2} + 276 q^{1/2}
+ 11202 q^{3/2} + \cdots~.$$

This theory enjoys a hidden $N=1$ supersymmetry.  At $c=12$, the dimension of the twist-field
which interpolates between the untwisted and twisted sectors of the orbifold is $\Delta = {3\over 2}$.  This is the
correct dimension for a supercharge, and there is in fact a $Co_0$-invariant choice of spin-3/2 field
whose OPEs closes on an $N=1$ superconformal algebra [\!\!\citenum{Duncan}].  The Ramond sector
of this theory has 24 ground states, transforming in the {\bf 24} irreducible representation of $Co_0$.

It is actually possible to view this theory as furnishing a realization of further enhanced chiral algebras.
This is a viewpoint explored in [\!\!\citenum{six,spinseven,DuncanMack}].  The basic idea is that one can construct
$U(1)$ or $SU(2)$ currents out of the free fermions $\psi_i$.  With appropriate choices, these can then
rotate the single supercharge of the $N=1$ superconformal theory to give additional supercharge(s)
filling out e.g. an $N=2$ or $N=4$ superconformal algebra.  Each choice of superalgebra gives rise to a 
moonshine connecting various mock modular forms (which arise as twining functions), to a preferred
finite group (arising as a subgroup of $Co_0$ which preserves suitable structure).\footnote{The
precise subgroups which give a moonshine, in the sense that all twining functions obey a suitable analogue of 
the genus zero property of the Monstrous case, are being investigated in [\!\!\citenum{FranSar}].}
  The various choices
explored in [\!\!\citenum{six,spinseven}]
are summarized in the table below.  In each case, one obtains a global symmetry group $G$ as a 
subgroup of $Co_0$ which fixes a $k$-plane in the ${\bf 24}$.  The new generators of
the corresponding superalgebra are constructed out of $k$ of the 24 free fermions.  It is natural to associate lattice geometries to each superalgebra, as the ${\bf 24}$ of $Co_0$ is canonically associated to the Leech lattice.

\begin{table}[htb]
\begin{center}
\begin{tabular}{c|c|c}
Superalgebra& Lattice geometry & Global symmetry group\\\hline $N=0$& $\mathbb R^{24}$& $Spin(24)$\\
$N=1$&$\Lambda_{Leech}$ & Conway$_0$\\
$Spin(7)$&$\Lambda_{Leech}$, fixed 1-plane & $M_{24}$\\
$N=2$&$\Lambda_{Leech}$, fixed 2-plane & $M_{23}$\\
$N=4$&$\Lambda_{Leech}$, fixed 3-plane & $M_{22}$\\
\end{tabular}\caption{The extended superconformal algebras of the Conway module, the global symmetry groups they preserve, and the natural objects these symmetry groups act on. We highlight the Mathieu groups for the extended superalgebras as these are (candidate) groups exhibiting moonshine. 
}\label{tbl:groups}
\end{center}
\end{table}

The options tabulated above correspond to $c=12$ realizations of these superalgebras, as one might
expect given a $c=12$ $N=1$ theory as the starting point.  
Here, we pursue a slightly different story, discussed in [\!\!\citenum{DuncanMack}] and [\!\!\citenum{K3inv}].  We can
find a $c=6$ realization of the $N=4$ superalgebra hidden here too.   Choose 4 real fermions
out of the 24, and construct from them two complex fermions $\psi^{\pm}_X$ and $\psi^{\pm}_Z$
in the obvious way (so e.g. $\psi_X^{\pm} = \psi_1 \pm i \psi_2$, and so forth).  Define the
currents
$$j^3 = {1\over 4} \left( \psi_X^- \psi_X^+ + \psi_Z^- \psi_Z^+ \right)$$
$$j^{\pm} = {i\over 2} \psi_X^{\pm} \psi_Z^{\pm} ~.$$
It turns out that $j^3, j^\pm$ generate an $SU(2)$ current algebra; and one can simply write down supercharges
promoting the $N=1$ supersymmetry to $N=4$, with this $SU(2)$ appearing as the
R-symmetry of the $N=4$ superconformal algebra.

Computing the Ramond sector partition function of the chiral CFT yields
$$Z^{s\natural}(\tau) = {\rm Tr}_R \left( (-1)^F q^{L_0 - {c\over 24}}\right) $$
$$ = {1\over 2} {1\over \eta^{12}(\tau)} \sum_{i=2}^{4} (-1)^{i+1} \theta_i^{12}(\tau,0) = 24~.$$
But we can flavor this with the $U(1)$ charges under the Cartan $U(1)$ in the $SU(2)$ we constructed
above.  And happily, it turns out that
$$Z^{s\natural}(\tau,z) \equiv {\rm Tr}_R\left( (-1)^{F} q^{L_0 - {c\over 24}} y^{J_0}\right)$$
$$ = {1\over 2} {1\over \eta^{12}(\tau)} \sum_{i=2}^{4} (-1)^{i+1} \theta_i^{2}(\tau,z) \theta_i^{10}(\tau,z)$$
(with $J_0$ being twice the zero mode of $j^3$)
is a weak Jacobi form of weight 0 and index 1, satisfying
$$Z^{s\natural}(\tau,0) = 24~.$$
This uniquely identifies it as the same weak Jacobi form appearing as the elliptic genus of the
K3 $\sigma$-model!  In the notation of Lecture 2, that is, 
$$Z^{s\natural}(\tau,z) = \phi_{K3}(\tau,z)~.$$

\subsection{Recovering and twining enumerative invariants}

From the fact that the Conway module ``knows about'' the elliptic genus of K3, we can use it to recover
the counting functions we studed in \S5.1-5.2.  

First of all, the Dijkgraaf-Moore-Verlinde-Verlinde formula [\!\!\citenum{DMVV}]
tells us that if we write the elliptic genus of K3 as
$$\phi_{K3}(\tau,z) = \sum_{n\geq 0, \ell} c(4n-\ell^2) q^n y^{\ell}$$
then

$$\sum_{n \geq 0} \phi_{K3^{[n]}}(\tau,z) p^{n-1} = 
\prod_{r,s,t \in {\bf Z}, r>0, s \geq 0} (1-q^s y^t p^r)^{-c(4rs-t^2)}~.$$
That is, the elliptic genera of all of the Hilbert schemes of $n$ points on K3, are determined by
the coefficients of the K3 genus itself.

But it is also true that the elliptic genus $\phi_M$ of a manifold $M$, specializes to the $\chi_y$ genus:
$${\rm lim}_{\tau \to i \infty} \phi_{M}(t,z) = y^{-d/2} \chi_{-y}(M)~.$$
It follows that from the elliptic genus of K3 alone, using the DMVV formula and this specialization,
we can recover the KKV formula (and hence also the Yau-Zaslow counting formula, by further
specialization).  The KKV invariants are therefore natural observables in our moonshine module.

In fact, we can also recover the doubly-graded KKP invariants.  In addition to the $U(1)$ current
$j^3$, there is another natural $U(1)$ we can write down after choosing $\psi^{\pm}_X$ and
$\psi^{\pm}_Z$, namely
$$k^3 = {1\over 4}\left(-\psi^-_X \psi^+_X + \psi^-_Z \psi^+_Z \right)~.$$
Then it is reasonable to compute
$$Z^{s\natural}(\tau,z,w) = {\rm Tr}\left(-(1)^F q^{L_0 - {c\over 24}} y^{J_0} u^{K_0}\right)$$
with $K_0$ being twice the zero mode of $k^3$.  It is easy to see
$$Z^{s\natural}(\tau,z,w) = {1\over 2} {1\over \eta^{12}(\tau)} \sum_{i=2}^4 
(-1)^{i+1} \theta_i(\tau,z-w) \theta_i(\tau,z+w) \theta_i^{10}(\tau,0)~.$$

A nice fact about the genera $Z^{s\natural}(\tau,z)$ and 
$Z^{s\natural}(\tau,z,w)$ is that, as they are defined as traces in the moonshine module,
they can naturally be twined by symmetries of the associated CFT.  However, we should be
careful to twine only by symmetries which preserve the structure of choices
$\psi^{\pm}_X$ and $\psi^{\pm}_Z$ -- that is, four-plane preserving elements of 
$Co_0$.  For any such $g$ in a four-plane preserving subgroup of $Co_0$, we can
define
$$Z_g^{s\natural}(\tau,z,w) = {\rm Tr}\left( g (-1)^F q^{L_0 - {c\over 24}} y^{J_0} u^{K_0}\right)~.$$
Specialization to $w=1$ yields the twined $Z^{s\natural}(\tau,z)$.

As the K3 CFT also enjoys symmetries under 4-plane preserving subgroups of $Co_0$, and
we can simply identify $Z^{s\natural}(\tau,z)$ with $\phi_{K3}(\tau,z)$, it is natural to conjecture
that the twined moonshine functions control also the twining genera of the K3 CFT.
This also applies to conjectural twined refined invariants of the K3 CFT.
Precise conjectures to this effect -- which are bit too involved to state here -- appear in [\!\!\citenum{K3inv}].  The interplay of this story with the
more detailed physical story involving 3D string theory, described in \S3, suggests that
the $Co_0$ module here is only relevant for twinings by symmetries of K3 that can be 
obtained by direct lift (without symmetry breaking) from the $Co_0$ point in the moduli
space of $K3 \times T^3$ compactifications.

It seems natural to mention at the close of this lecture that connections between moonshine,
enumerative geometry of Calabi-Yau threefolds, and heterotic strings on $K3 \times T^2$ have 
appeared in [\!\!\citenum{CYthree,Luest}].  The results of [\!\!\citenum{KlemmKatz}] and references therein promise a rich, rapidly evolving story of modular
forms associated to topological strings on Calabi-Yau threefolds, and we expect many further
developments along the lines of this lecture in the future.

\medskip
\centerline{\bf{Acknowledgements}}
\medskip
I thank the organizers of the 18th International Congress on Mathematical Physics for their hospitality in
Santiago de Chile in July of 2015, and the organizers and students of the ``School on Mathematics of
String Theory" for providing stimulating discussions at the Centre International de Recontres Mathematiques
in Marseille in April of 2016.  I am also very grateful to my collaborators -- especially N. Benjamin, M. Cheng, J. Duncan, E. Dyer,
A. Fitzpatrick, S. Harrison, N. Paquette, and R. Volpato -- for teaching me a great deal about moonshine, extremal gravity,
and the mathematics of modular forms.  I thank K. Becker, J. Duncan, S. Harrison, N. Paquette and T. Wrase for
comments on the manuscript.  Finally, I acknowledge significant help from N. Benjamin with the figures.

\newpage




\end{document}